\documentclass[11pt] {article}
\usepackage[a4paper,margin=1in]{geometry}
\usepackage{lineno}

\usepackage[T1]{fontenc} %
\usepackage[normalem]{ulem} %

\usepackage[usenames,dvipsnames,svgnames,x11names]{xcolor}

\usepackage{cite}

\usepackage{balance}

\usepackage{listings}

\lstdefinelanguage{XML}
{
basicstyle=\ttfamily\footnotesize,
  morestring=[b]",
  moredelim=[s][\bfseries\color{Maroon}]{<}{\ },
  moredelim=[s][\bfseries\color{Maroon}]{</}{>},
  moredelim=[l][\bfseries\color{Maroon}]{/>},
  moredelim=[l][\bfseries\color{Maroon}]{>},
  morecomment=[s]{<?}{?>},
  morecomment=[s]{<!--}{-->},
  commentstyle=\color{gray},
  stringstyle=\color{blue},
  identifierstyle=\color{red}
}

\usepackage{moreverb}

\usepackage[nounderscore]{syntax}

\usepackage[pdftex]{graphicx}
\graphicspath{{./figures/}}
\DeclareGraphicsExtensions{.pdf}

\usepackage[cmex10]{amsmath}
\usepackage{amssymb}
\usepackage{mathtools}
\usepackage{amsthm}
\usepackage{amsfonts}
\usepackage{gensymb}

\usepackage{subfig} %

\usepackage{algorithmicx}
\usepackage{algpseudocode}
\usepackage[ruled]{algorithm}
\definecolor{light-gray}{gray}{0.75}
\algrenewcommand{\algorithmiccomment}[1]{\hskip3em{{\footnotesize \textcolor{light-gray}{$\blacktriangleright$}}} #1}

\usepackage{multirow} %
\usepackage{rotating} %
\usepackage{booktabs} %
\usepackage{colortbl} %
\usepackage{tablefootnote} %

\usepackage{array}
\newcolumntype{L}[1]{>{\raggedright\let\newline\\\arraybackslash\hspace{0pt}}m{#1}}
\newcolumntype{C}[1]{>{\centering\let\newline\\\arraybackslash\hspace{0pt}}m{#1}}
\newcolumntype{R}[1]{>{\raggedleft\let\newline\\\arraybackslash\hspace{0pt}}m{#1}}

\usepackage[pdftex,colorlinks=true,urlcolor=blue,citecolor=blue]{hyperref}

\usepackage{xspace}

\usepackage{enumitem}

\usepackage{blindtext}

\newcommand{\rp}{\texttt{RIPPLE}\xspace}
\newcommand{\rpp}{\texttt{RIPPLE++}\xspace}

\newcommand{\drc}{\textbf{DRC}\xspace}
\newcommand{\drg}{\textbf{DRG}\xspace}
\newcommand{\dnc}{\textbf{DNC}\xspace}
\newcommand{\dng}{\textbf{DNG}\xspace}
\newcommand{\rc}{\textbf{RC}\xspace}
\newcommand{\K}{\textsc{K}}
\newcommand{\M}{\textsc{M}}
\newcommand{\B}{\textsc{B}}

\def\orcid#1{\kern .08em\href{https://orcid.org/#1}{\includegraphics[keepaspectratio,width=0.7em]{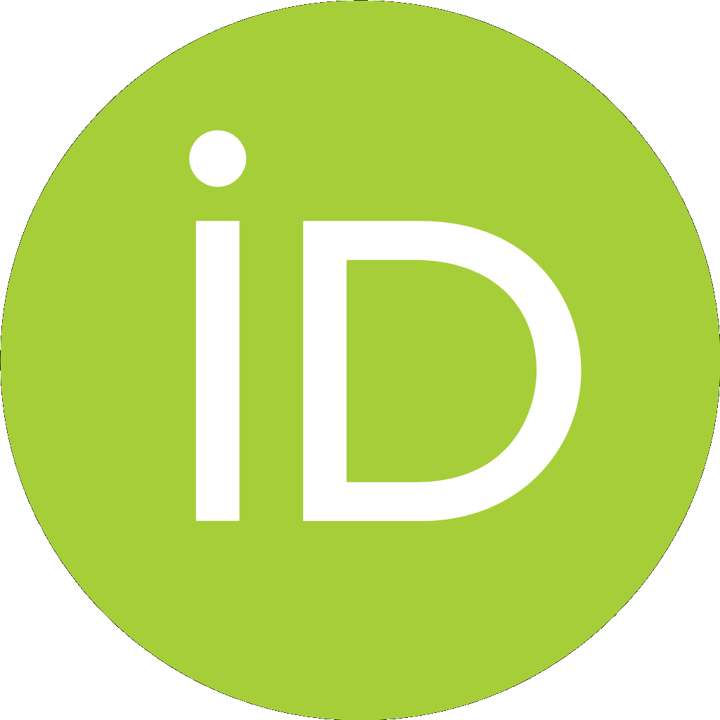}}}

\begin{document}

\title{\rpp : An Incremental  Framework for Efficient GNN Inference on Evolving Graphs~\thanks{~Extended full-length version of paper that appeared at ICDCS 2025: \textit{``\rp: Scalable Incremental GNN Inferencing on Large Streaming Graphs'', Pranjal Naman and Yogesh Simmhan, in International Conference on Distributed Computing Systems (ICDCS), 2025}. DOI: \url{https://doi.org/10.1109/icdcs63083.2025.00088}}
}

\author{Pranjal Naman\orcid{0009-0000-9912-9522}, Parv Agarwal\orcid{0009-0004-5839-6471}, Hrishikesh Haritas\orcid{0009-0001-3931-2699} and Yogesh Simmhan$^1$\orcid{0000-0003-4140-7774}\\~\\
\em Department of Computational and Data Sciences (CDS),\\
\em Indian Institute of Science (IISc),\\
\em Bangalore 560012 India\\~\\
\texttt{Email:\{pranjalnaman, simmhan\}@iisc.ac.in}
}
\date{}
\maketitle

\begin{abstract}
Real-world graphs are dynamic, with frequent updates to their structure and features due to evolving vertex and edge properties. These continual changes pose significant challenges for efficient inference in graph neural networks~(GNNs). Existing \textit{vertex-wise} and \textit{layer-wise} inference approaches are ill-suited for dynamic graphs, as they incur redundant computations, large neighborhood traversals, and high communication costs, especially in distributed settings. Additionally, while sampling-based approaches can be adopted to approximate final layer embeddings, these are often not preferred in critical applications due to their non-determinism.
These limitations hinder low-latency inference required in real-time applications. To address this, we propose \rpp, a framework for streaming GNN inference that efficiently and accurately updates embeddings in response to changes in the graph structure or features. 
\rpp introduces a generalized incremental programming model that captures the semantics of GNN aggregation functions and incrementally propagates updates to affected neighborhoods. \rpp accommodates all common graph updates, including vertex/edge addition/deletions and vertex feature updates.
\rpp supports both single-machine and distributed deployments. On a single machine, it achieves up to $56\K$ updates/sec on sparse graphs like Arxiv~($169\K$ vertices,  $1.2\M$ edges), and about $7.6\K$ updates/sec on denser graphs like Products~($2.5\M$ vertices, $123.7\M$ edges), with latencies of $0.06$--$960$ms, and outperforming state-of-the-art baselines by $2.2$--$24\times$ on throughput.
In distributed settings, \rpp offers up to $\approx25\times$ higher throughput and $20\times$ lower communication costs compared to recomputing baselines.
\end{abstract}

\section{Introduction}

Graph Neural Networks~(GNNs)~\cite{kipf2016semisupervised} have the ability to learn low-dimensional representations that capture both the topology and attributes of graph datasets. Unlike traditional neural networks that assume grid-like inputs (e.g., images or sequences), GNNs can capture both the vertex/edge features and the topological structure by aggregating and transforming a vertex's multi-hop neighbors into embeddings for training.
This ability to learn expressive 
features in the property graph, and use them for inference tasks such as vertex labeling, subgraph labeling, link prediction, etc., makes GNNs a powerful tool. It is used in wide-ranging linked data applications: for detecting financial frauds~\cite{liu2021pick, dou2020enhancing}, predicting traffic flow~\cite{guo2019traffic}, analyzing social networks~\cite{yang2021gnnsocial, kipf2016semisupervised}, and making e-commerce recommendations~\cite{chen2020tgcn, chang2021sequential, wu2022gnnrec, lyu2020gnnrec2}.

Such applications frequently operate on \textit{large-scale graphs}, consisting of millions to billions of vertices and edges, e.g., financial platforms monitor millions of real-time transactions between users, and urban traffic systems manage sensor data from hundreds of thousands of junctions. 
These settings often impose real-time latency requirements during GNN   inference~\cite{wu2025inkstream,lin2022platogl,zhang2023graphagile, wang2021apan}. 
For instance, a delay in identifying suspicious transactions in financial networks can allow fraudulent activities to occur, leading to financial loss. Similarly, delays in real-time traffic flow prediction used for traffic signal optimization can exacerbate congestion during peak hours. 

\begin{figure}[t]
    \centering
    \includegraphics[width=0.8\columnwidth]{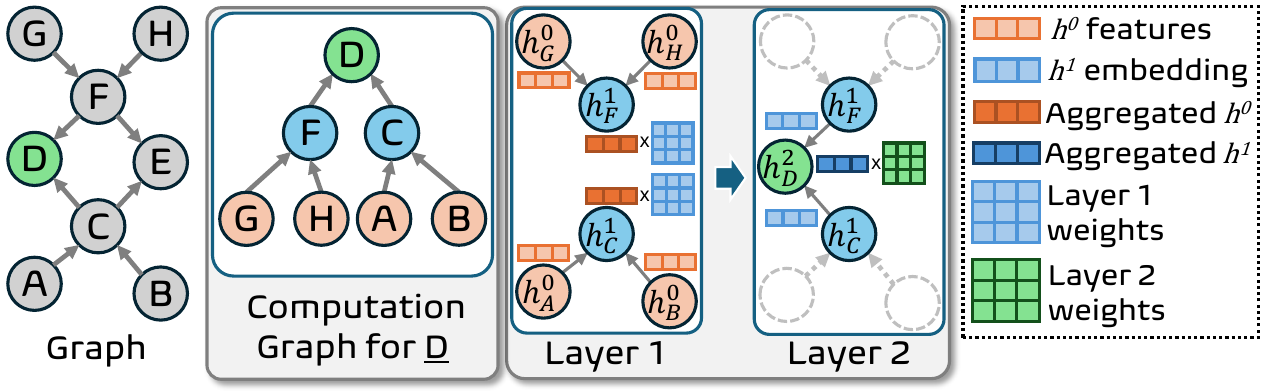}
    \caption{GNN inference on static graphs for vertex $D$. $h^2_D$ is the final layer embedding that maps to a predicted label.}
    \vspace{-0.1in}
    \label{fig:static-inference-1}
\end{figure}


\subsection{GNN Inference on Static Graphs}
GNN inference involves performing a forward pass through the trained Neural Network (NN) model. When, say, predicting the label for vertex $D$ in Fig.~\ref{fig:static-inference-1}, we compute its $L$-hop in-neighbors~(called the \textit{computational graph}), and then propagate the aggregated leaf features to the input layer of the $L$-layer NN to generate embeddings for the penultimate level of the computation graph. We pass these embeddings higher up the tree as we pass through each NN layer. The final prediction for the vertex is the final-layer embedding produced by this forward pass.

When inference has to be done with low latency, the simplest approach is to perform a forward pass once on all graph entities using the trained GNN model, cache the predicted results, and return them rapidly using a simple lookup~\cite{yin2022dgi}.

\begin{figure}[t]
    \centering
    \includegraphics[width=0.78\columnwidth]{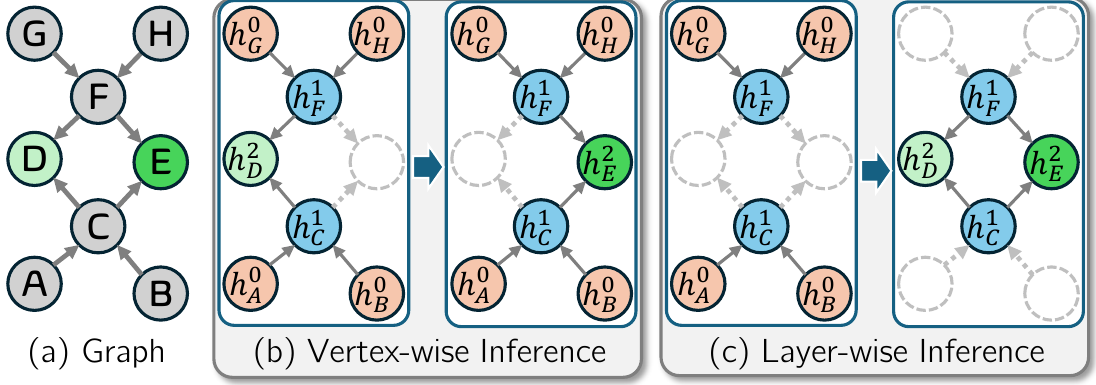}
    \caption{\textit{Vertex-wise} vs. \textit{Layer-wise} inference on static graphs.}
    \vspace{-0.1in}
    \label{fig:static-inference-2}
\end{figure}

There are two approaches for generating the final layer embeddings in a static graph.
\textit{Vertex-wise} inference~(Fig.~\ref{fig:static-inference-2}, center) 
is akin to just performing the forward pass of GNN training (training also includes the backward pass to update weights).
The embeddings within an $L$-hop neighborhood of each target vertex~(\textit{green}) being labeled are aggregated, assuming a vertex labeling GNN task. During training, the neighborhood of the computational graph is randomly sampled to keep the subgraph size manageable and still achieve good model accuracy~\cite{hamilton2017sage}. However, such sampling affects the correctness and \textit{deterministic} nature of the predictions if done during inference~\cite{kaler2022accelerating}, making it unsuitable for many real-world applications due to both \textit{non-determinism and approximation},
e.g., the same transaction should be classified as a fraud across multiple inference requests, in the absence of other changes. Fig.~\ref{subfig:inf-fanout-spread} shows this trade-off for the Reddit social network graph~($233\K$ vertices, $114\M$ edges; experiment setup
described in \S~\ref{subsec:eval-expsetup}). As the sampling fanout size increases, we get a better and more predictable accuracy~(\textit{left y-axis}, box plots) but also a higher per-vertex average inference time~(\textit{right y-axis}, \textit{red} marker).
Guaranteeing both deterministic and accurate predictions requires the entire $L$-hop neighborhood to be considered during inference, which can lead to \textit{neighborhood explosion}~\cite{hamilton2017sage}, with significantly higher memory and computational demands. 

So, a \textit{layer-wise} inference approach (Fig.~\ref{fig:static-inference-2}, right) is typically employed for bulk processing~\cite{yin2022dgi}. Here, the embeddings for each hop~(colored vertex layers) are computed for all vertices in the graph and used as inputs to calculate the embeddings for the next layer~(\textit{orange} to \textit{blue} to \textit{green}). This avoids the explosion in computational graph sizes by leveraging overlaps in vertices at the same layer with the neighborhoods of proximate vertices, and prevents redundant computation, which occurs in vertex-wise inference, e.g., $h^1_F$ and $h^1_C$ are recomputed for inference of $D$ and $E$ but computed only once in layer-wise.


\subsection{GNN Inference on Evolving Graphs} 
GNN inference becomes even more complex when operating on graphs that \textit{evolve continuously} in the presence of vertex/edge additions/deletions, or changes to their features~\cite{jie2025helios, besta2021practice}. E.g., a friend being added or removed in a social network causes an edge addition/deletion~\cite{bai2020temporal}; changing traffic flows in a road network updates an edge feature~\cite{guo2019traffic}, transactions in a fintech network change the account-balance vertex feature~\cite{song2023financial}. Thousands of such dynamic graph mutations can occur per second, affecting the output of GNN predictions for entities in their neighborhood~\cite{wu2020deltagrad, mahadevan2023cost, ammar2016techniques}. However, performing \textit{layer-wise} inference on the \textit{whole graph} every time an update~(or a small batch of updates) occurs is too time-consuming for latency-sensitive applications. Instead, we can limit the recomputation of the vertex labels only to those that fall within the $L$-hop out-neighborhood~(\textit{affected neighborhood}) of entities that have been updated in the graph.

\begin{figure}[t]
\vspace{-0.15in}
  \centering
  \begin{minipage}[t]{0.35\columnwidth}
    \centering
    \includegraphics[width=\textwidth]{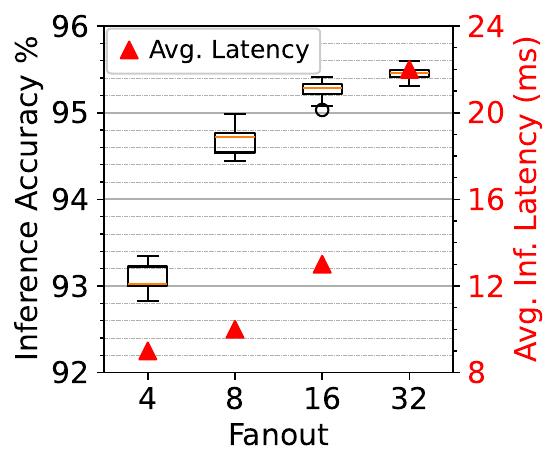}
    \caption{Effect of neighborhood sampling on vertex-wise inference accuracy and latency (Reddit graph, 3-layer SAGEConv).}
    \label{subfig:inf-fanout-spread}
  \end{minipage}\qquad\qquad
  \begin{minipage}[t]{0.44\columnwidth}
    \centering
    \includegraphics[width=\textwidth]{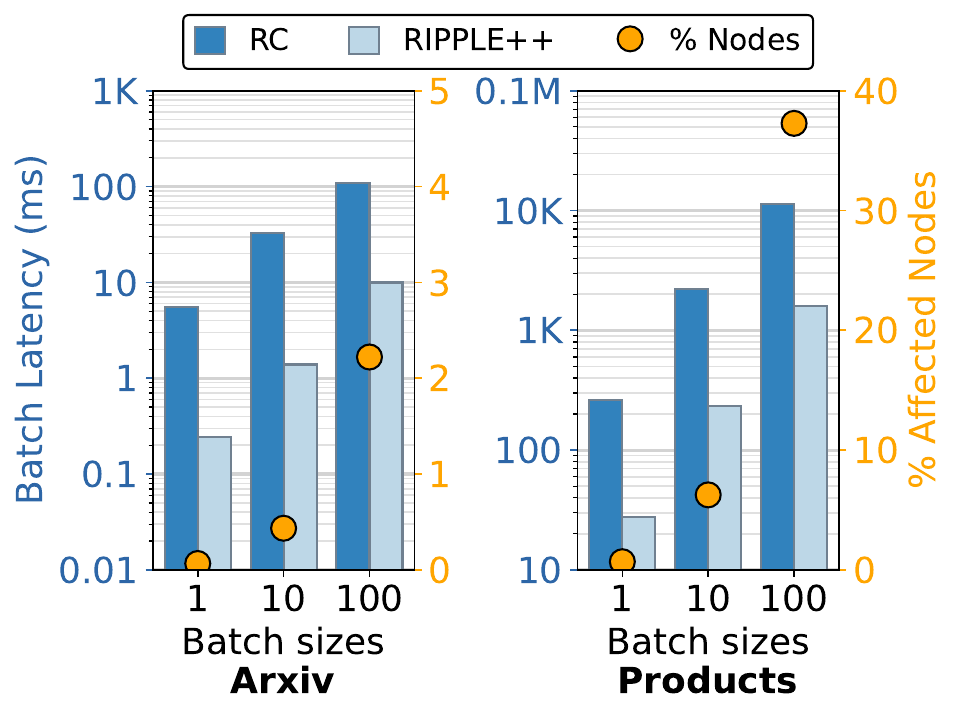}
    \caption{\% of affected vertices and inference latency per batch-update, for \rpp~(RP) and baseline RC, with differing batch sizes (3-layer GraphConv).}
    \label{subfig:affected-nodes}
  \end{minipage}
  \vspace{-0.1in}
\end{figure}

Notably, the extent of these cascading effects grows with the average degree of the graph and the number of updates applied in a batch, as more entities fall within the affected neighborhood of the modified vertices. E.g., in Fig.~\ref{subfig:affected-nodes}, the fraction of vertices affected for the \textit{Arxiv} graph~(average vertex degree of $6.9$) increases from $0.06\%$ to $2.2\%$ of the total vertices, as the batch size~($bs$) increases from $1$ to $100$. However, for the \textit{Products} graph~(degree $50.5$), the growth is sharper, from $0.7\%$ to $\approx37\%$. This also directly impacts the \textit{latency} to translate incoming graph updates into updated GNN embeddings and inversely affects the \textit{update throughput} that can be supported.
As the affected neighborhood size increases, so does the per-batch latency to perform \textit{inference recomputations}~(\rc) upon updates, growing from $5.5$ms to $\approx110$ms for \textit{Arxiv} using $bs$ from $1$--$100$, and from $\approx260$ms to $11.3$s for \textit{Products}. The throughput of updates processed to give fresh GNN predictions is $177$--$916$~up/s for Arxiv but only $3.7$--$8.8$~up/s for Products.

\subsection{Large Memory Requirements of Layer-wise Inference} Layer-wise GNN inference requires the whole graph and its embeddings to be present in memory\footnote{Interestingly, layer-wise GNN inference does not benefit much from using GPUs since the computational load is modest~(Fig.~\ref{fig:gpu-analysis} in Appendix). Instead, computation on CPUs offers comparable performance and also benefits from the available larger host memory. 
}. But graphs such as \textit{Papers}~($111\M$ vertices, $1.6\B$ edges, $128$ features)~\cite{hu2020open} and their embeddings can easily take $>128$~GiB of RAM. Another significant challenge in handling real-world workloads is the immense scale of the graph data. As the graph sizes grow, managing the inference process on a single machine becomes more challenging due to the increased memory demands. This necessitates efficient \textit{distributed execution} of incremental computation on a compute cluster to support larger graphs. 

Recent works like InkStream~\cite{wu2025inkstream}, that also perform incremental GNN inference like us, only support graph edge updates, work with fewer GNN architectures, and lack a distributed execution model for scaling to larger graphs.

\subsection{Contributions}
In this paper, we present \rpp, a low-latency framework for streaming GNN inference that efficiently handles real-time updates to large-scale graphs using incremental computation. It uniquely applies a delta to undo previous embedding aggregations, and redoes them using updated embeddings to sharply reduce any recomputation. \rpp offers both a single-machine and a distributed execution model. Our technique \textit{avoids redundant computation} of embeddings for the entire neighborhood of updates and instead, scopes it to just a subset of the operations.
In particular, when one~(or a few) vertices ($V'$) among the in-neighbors ($V$) of another vertex ($u$) get updates, rather than aggregate the embeddings from all $V$ vertices, we \textit{incrementally aggregate} only the deltas from the updated vertices $V'$. This reduces the number of computations from $k=|V|$ to $k'=|V'|$, which is often an order of magnitude smaller.
Further, we support diverse mutations such as vertex/edge/feature updates and GNNs with linear, non-linear, and attention-based aggregation, while guaranteeing deterministic and accurate behavior, along with a distributed execution model for large graphs, unlike SOTA approaches~\cite{wu2025inkstream} that only support edge-level updates and on only a single machine.

Specifically, we make the following contributions:

\begin{enumerate}[leftmargin=*]
    \item We propose \rpp, a scalable incremental GNN inference framework that supports streaming edge and vertex additions and deletions, and feature updates, for GNN models that can include \textit{monotonic} and \textit{accumulative} aggregators, and also \textit{attention-based} architectures~(\S~\ref{sec:design}).

    \item We present a detailed analytical comparison of our incremental approach~(\rpp) against full recomputation~(\rc), characterizing the conditions under which incremental inferencing offers best benefits~(\S~\ref{subsubsec:benefits-analysis}).

    \item We design a locality-aware routing strategy to place new vertices, which minimizes edge cuts and improves distributed execution under graph evolution compared to hash-based routing~(\S~\ref{subsubsec:dist-placement}).

    \item We perform detailed experiments of \rpp on four real-world graphs with $169\K$--$111\M$ vertices and $1.2\M$--$1.6\B$ edges, for seven workloads across four GNN models and aggregation functions. We also compare it against other recomputation strategies and a state-of-the-art (SOTA) framework, InkStream~\cite{wu2025inkstream}.
    \rpp achieves up to $56\K$~up/s inference throughput on a single-machine that is $2.2$--$24\times$ faster than InkStream, and 
    $25\times$ improvement over the recompute baseline
    in a distributed setting~(\S~\ref{sec:evaluation}).
\end{enumerate}
We also provide background on GNN training and inference (\S~\ref{sec:background}), contrast \rpp with related works (\S~\ref{sec:related}), and conclude with a discussion of future directions~(\S~\ref{sec:conclusion}).

In a prior conference paper~\cite{naman2025ripple}, we introduced the key ideas of incremental GNN inference in \rp. 
This article significantly extends it as \rpp: (1) 
We generalize the incremental technique to support complex GNN architectures with monotonic operators (e.g., max) and graph-attention (e.g., GAT), beyond just \textit{accumulative} operators supported earlier;
(2) We extend the design to support vertex additions and deletions,
to complement edge/vertex feature updates supported earlier;
(3) We offer a detailed analysis of incremental inferencing against full-recompute, identifying the scenarios where we offer the most benefit;
(4) We propose a locality-aware update routing to reduce edge-cuts and improve load balancing in a distributed setting, which goes beyond our earlier hash-based routing; 
and
(5) We substantially expand the experiments 
with two new hybrid incremental GNNs, study the effect of number of GNN layers,
scale the evaluation to $20\M$ updates, 
and compare against the SOTA InkStream work~\cite{wu2025inkstream}.
These make \rpp a general and scalable platform for low-latency GNN inference on large evolving graphs.


\section{Background}\label{sec:background}

\begin{figure}[t]
    \centering
    \includegraphics[width=0.7\columnwidth]{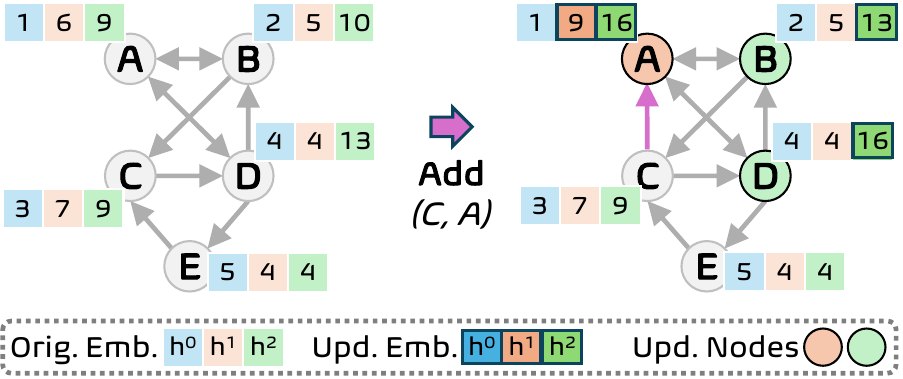}
    \caption{Cascading effect on the vertex embeddings due to an edge addition during 2-layer GNN inference.}
    \vspace{-0.1in}
    \label{fig:streaming-edge-add}
\end{figure}

\subsection{Training vs. Inference of Graph Neural Networks}\label{subsec: background-train-v-inf}

GNNs are trained using iterative forward and backward passes, similar to Deep Neural Networks (DNNs). During the \textit{forward pass}, when training over a labeled vertex $u$, the layer-\textit{l} of an \textit{L}-layer GNN uses the \textsc{Aggregate} function to accumulate the $(l - 1)$ embeddings of the neighbors $\mathcal{N}(u)$ of the vertex (Fig.~\ref{fig:static-inference-1}). The aggregated embedding vector $x^l_u$ is then processed by an \textsc{Update} function, which is the learnable component of the network, and passed through a nonlinear function $\sigma$. This repeats $L$-times, once per layer, to get the final layer embedding $h^L$ for the training vertex. This ends the forward pass. We then do a \textit{backward pass} across $L$ layers to update the learnable parameters using a loss function.
\begin{align}
    x^l_u =&~ \textsc{Aggregate}^l(\{h^{l-1}_v, v \in \mathcal{N}(u)\}) \label{eq:gnn}\\
    h^l_u =&~ \sigma(\textsc{Update}^l(h^{l-1}_u, x^l_u))
\end{align}
During GNN inference, only the forward pass is performed to compute the final layer embedding for the target vertex, which maps to the predicted label.
For static graphs, this can be 
pre-calculated, thus avoiding compute at inference time.

\subsection{Inference on Streaming Graphs}\label{subsec:bg-inf-on-streaming}
Unlike for static graphs, pre-calculating the final layer embeddings and labels for all vertices is not beneficial for dynamic graphs. Each update to a vertex or edge triggers a cascade of embedding updates across multiple vertices at each hop, ultimately changing the embeddings and, consequently, the predicted labels of all vertices at the final hop. 

Fig.~\ref{fig:streaming-edge-add} shows a unit-weighted graph with pre-calculated embeddings for a $2$-layer GNN with $sum$ as the aggregator. When edge $(C, A)$ is added, the $h^1_A$ and $h^2_A$ embeddings of $A$ get updated, which cause cascading updates to the $h^2$ embeddings of $\{B, D\}$, possibly changing the predicted labels of $\{A, B, D\}$ in the final hop. Notably, the embeddings of $C$ and $E$ remain unaffected. Any change in the graph topology or the vertex features only \underline{\textit{ripples}} through to a maximum of $L$-hops, for an $L$-layer GNN. So, it is not necessary to update the embeddings of \textit{all} vertices.
The fraction of affected vertices per batch can span from $<3\%$ of vertices for \textit{Arxiv} to $\approx 37\%$ of vertices for \textit{Products} (Fig.~\ref{subfig:affected-nodes}). At the same time, it is crucial to propagate the effect of the dynamic graph updates quickly and accurately for timely inference.

We can classify the inference of streaming graphs into trigger-based and request-based. \textit{Trigger-based} inference immediately notifies the application of any changes to the predicted vertex label due to graph updates. In contrast, \textit{request-based} inference follows a pull-based model, where the application queries the label of a specific entity in the graph as needed. Both classes have unique requirements and different strategies, e.g., update propagation can be done lazily for rarely accessed vertices in the graph for the request-based inference model. In this paper, we tackle trigger-based inference.

\begin{figure}[t]
    \centering
    \includegraphics[width=0.8\columnwidth]{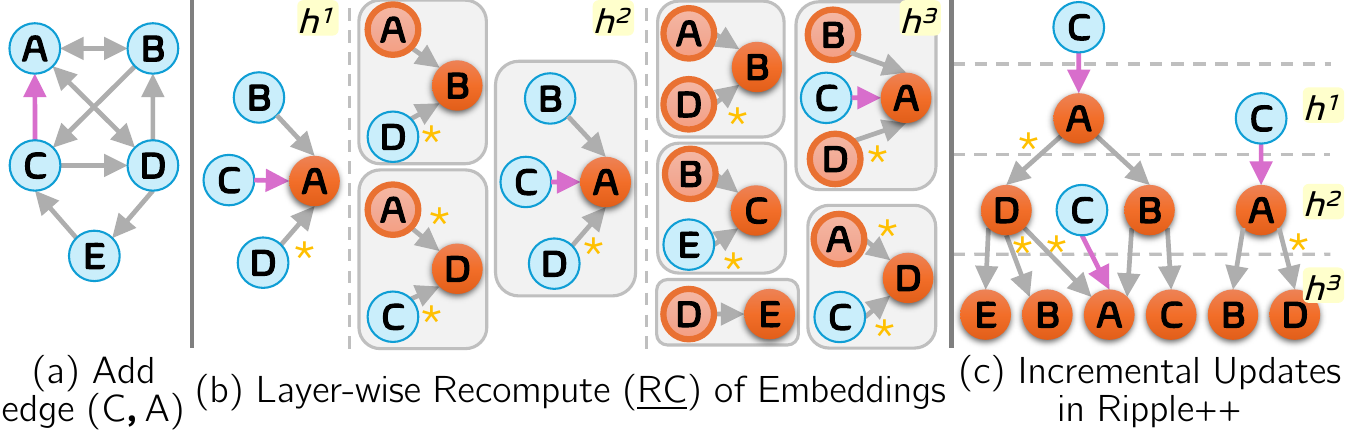}
        \caption{Recomputation and \rpp's Incremental Comp. on Edge addition. Dark orange vertices update their embeddings in a hop (3-layer GNN). Edges with \textcolor{orange}{`$*$'} indicate remote messages passed between machines during distributed execution across partitions $\{A,B,C\}$ and $\{D, E\}$, discussed in \S~\ref{sec:design:distr}.}
        \vspace{-0.1in}
        \label{fig:ripple-single-schematic}
\end{figure}

\subsection{Aggregation Functions}\label{subsec:bg-agg-funcs}
In a multi-layered GNN, the \textit{neighborhood aggregation} forms a key operation, where each vertex combines the information from its neighbors to generate its representation using Eqn.~\ref{eq:gnn}. Existing research on GNNs often uses \textit{linear accumulative} functions like $sum$, $mean$, or $weighted~sum$ based on edge weights or edge attention; occasionally, they consider \textit{non-linear monotonic} aggregation functions like $max$ or $min$. 
Kipf et al.~\cite{kipf2016semisupervised} propose the Graph Convolution Network~(GCN) architecture, which adopts a weighted summation aggregation, while GraphSAGE \cite{hamilton2017sage} typically uses sum or mean. Graph Attention Networks~(GAT)~\cite{velivckovic2017graph} follow an attention-based summation where the attention on an edge is calculated based on the features of the participating vertices. Xu et al.~\cite{xu2018powerful} show that the expressiveness of the sum aggregator overshadows other functions, and use this in Graph Isomorphism Networks~(GIN). Dehmamy et al.~\cite{dehmamy2019understanding} and Corso et al.~\cite{corso2020principal} show that the expressive power of a GNN module can be increased by using a combination of aggregators, e.g., the former uses a combination of sum and mean, while the latter uses both linear and non-linear aggregators. 
With this in mind, we support a variety of such aggregation functions across multiple GNN architectures~(Table~\ref{tab:gnn_aggregation}).

\begin{table}
	\centering
	\caption{Popular Aggregation functions in GNNs.}
	\label{tab:gnn_aggregation}
	\begin{tabular}{l|l|l}
		\hline
		\textbf{Aggregation Fn.} & \textbf{Definition}                                                                   & \textbf{Used By}                                                                                       \\ \hline\hline
		$sum$                    & \(\mathbf{h}_i = \sum_{j \in \mathcal{N}(i)} \mathbf{h}_j\)                            & \cite{hamilton2017sage}\cite{xu2018powerful}\cite{dehmamy2019understanding} \\ \hline
		$mean$                   & \(\mathbf{h}_i = \frac{1}{|\mathcal{N}(i)|} \sum_{j \in \mathcal{N}(i)} \mathbf{h}_j\) & \cite{hamilton2017sage}\cite{dehmamy2019understanding}                                                                                                                       \\ \hline
            $max/min$                   & \(\mathbf{h}_i = max_{j \in \mathcal{N}(i)} \mathbf{h}_j\) & \cite{hamilton2017sage}\cite{dehmamy2019understanding}                                                 \\ \hline
		$weighted~sum$          & \(\mathbf{h}_i = \sum_{j \in \mathcal{N}(i)} \alpha_{ij} \cdot \mathbf{h}_j\)          & \cite{kipf2016semisupervised}
         \\ \hline
		$attention$          & \(
\mathbf{h}_i = \sum_{j \in \mathcal{N}(i)} \alpha(\mathbf{h}_i,\mathbf{h}_j) \cdot \mathbf{h}_j
\)
         & \cite{velivckovic2017graph} \\
        \hline
	\end{tabular}
    \vspace{-0.1in}
\end{table}


\section{\rpp System Design}\label{sec:design}
In this section, we present the design for \rpp, a GNN inference framework for streaming graphs that supports \textit{trigger-based} applications that need to be notified of changes to the predictions of any graph entity upon receiving updates as soon as possible. \rpp uses incremental computation over a batch of updates, intelligently using the delta in prior embeddings to avoid redundant computation of parts of the computation neighborhood. \rpp supports both an efficient single-machine execution if the entire graph, its features and embeddings fit in the RAM of a single server, and a distributed execution to scale to larger graphs with efficient communication primitives.

\subsection{Preliminaries}\label{subsec:assumptions}
We make certain simplifying assumptions when designing \rpp. The GNN model used for inference has $L$ layers. As bootstrap, the initial embeddings have been calculated for all existing graph entities prior to new updates streaming in. Each vertex $u \in V$ has its features $h_u^0$ and intermediate and final layer embeddings $h_u^l$~($l \in [1, L]$). The initial embedding matrices,  $H_{T_0}^{l}$ of layer $l$, are generated using the GNN model and stored, and the current labels of all vertices can be extracted from $H_{T_0}^{L} $. These serve as starting points when new updates arrive. 

\rpp is designed for GNN model architectures that support either \textit{linear accumulative}, \textit{non-linear monotonic}, or attention-based aggregation functions~(Table~\ref{tab:gnn_aggregation}). \rpp works in one of two incremental modes -- \textit{completely incremental} or \textit{hybrid incremental}, which we describe later.
We use a GNN model for vertex classification as the running example, though this can be extended to other vertex- or edge-based tasks. \rpp supports five types of updates: \textit{edge additions}, \textit{edge deletions}, \textit{vertex additions}, \textit{vertex deletions}, and \textit{vertex feature changes}.

We assume that updates arrive \textit{continuously} and are \textit{batched} into fixed batch sizes~($bs$) that are then applied to the graph, triggering~(incremental) computation of the affected entities. The updated predictions are immediately made available to the consumers. Since we assume a high update rate of $100$--$1000$ updates per second~(up/s), this bulk operation amortizes the overheads, reduces redundant computation, and achieves higher throughput. The size of the batch is a hyperparameter that can be tuned to trade-off update throughput against batch execution latency. We evaluate \rpp for different batch sizes in our experiments, which in the future can guide the selection of dynamic batch sizes.


\subsection{Layer-wise Update Propagation through Recomputation}\label{subsec:design-rc}

We first describe a competitive baseline approach that performs \textit{layer-wise recomputing (RC)}, scoped to the neighborhood of updates.
When an edge/vertex/feature is updated, it causes cascading updates, starting from the \textit{root vertex} on which the update is incident and on the embeddings within the $L$-hop neighborhood~(Fig.~\ref{fig:streaming-edge-add}). 

An \textit{edge addition} $(u, v)$ immediately alters the $h^1_v$ embedding of the sink vertex $v$, which leads to further downstream changes. E.g., in Fig.~\ref{fig:ripple-single-schematic}(b), adding the edge $C \rightarrow A$ initially changes only the $h^1_A$ embeddings of sink vertex $A$ at hop distance $1$ from the root vertex $C$. But recomputing $h^1_A$ requires fetching the $h_0$ embeddings of \textit{all} the in-neighbors of $A$: $h^0_B$, $h^0_C$ and $h^0_D$, and aggregating them, followed by the update operation.

Similarly, for the next layer of updates to $h^2$ embeddings, the update to $h^1_A$ cascades to all out-neighbors $\{B, D\}$ for $A$, causing $h^2_B$ and $h^2_D$ to be updated by aggregating the new value of $h^1_A$ and prior value of $h^1_D$ for the neighbor $B$, and the new value of $h^1_A$ and prior value of $h^1_C$ for neighbor $D$.
In addition, the addition of the $C \rightarrow A$ edge will also cause the $h^2_A$ embeddings for $A$ to get updated, aggregating the new edge embedding from $h^1_C$ and prior values of $h^1_B$ and $h^1_D$. 
Hence, recomputation of the $h^2$ embeddings requires pulling the $h^1$ embeddings of all the in-neighbors of the affected vertices, $A, B$, and $D$. Similarly, the $h^3$ embeddings of the out-neighbors of $\{A, B, C, D, E\}$ will be updated by pulling the $h^2$ embeddings of their neighbors and recomputing the aggregation function over all of them, followed by the update function. These cascading changes are illustrated by the orange vertices in Fig.~\ref{fig:ripple-single-schematic}(b) at each layer.

An \textit{edge deletion} will traverse from the source vertex as the \textit{root}, and affect the same set of vertices at each hop as above. It requires similar recomputations, except now there will be \textit{one less neighbor} when updating $h^l_v$. The embedding update to $h^1_v$ will cause a similar cascading set of changes to downstream vertices. E.g., if edge $C \rightarrow A$ is removed at a later time, the only vertex whose $h^1$ embedding changes is $A$, followed by a similar propagation as when the edge was added.

A \textit{vertex addition} in itself should not trigger any updates, except to update its own embeddings. However, vertices can be created as a part of an edge addition. So, adding a vertex $u$ can trigger the same traversal from root $u$ as when an edge $(u, v)$ is added, if the sink vertex $v$ already exists but the source vertex $u$ needs to be created. But, if $u$ is the sink vertex of an added edge, only its own embedding is affected, as no downstream vertices exist to traverse to.

A \textit{vertex deletion} causes a delete of all edges incident upon. Further, with this vertex as root, it cascades along the entire deleted out-neighborhood, up to $L$ hops, potentially impacting a large neighborhood and being the most expensive of all update types. 
E.g., in Fig.~\ref{fig:ripple-single-schematic}(b), deletion of $C$ would first lead to the removal of edges $C \rightarrow A$, $C \rightarrow D$, $E \rightarrow C$, and $B \rightarrow C$, and then the deletion of the vertex $C$ itself. This is followed by an update propagation to $A$ and $D$, since these were downstream from $C$, and so on.

Lastly, when a \textit{vertex feature} is updated, the out-neighborhood that is affected is similar to vertex deletion, as
updating the feature $h^0_u$ for vertex $u$ impacts all its out-neighbors, and this impact cascades up to $L$ hops. E.g., in Fig.~\ref{fig:ripple-single-schematic}(b), after the edge addition $C \rightarrow A$, if the features of vertex $C$ change, it impacts its out-neighbors, $A$ and $D$, and leads to a larger propagation downstream. 

A key limitation of layer-wise recomputation is that updating the $h^l_v$ embedding of vertex $v$ at layer $l$ requires us to fetch the embeddings for the previous layer $h^{l-1}_u$ of \textit{all its in-neighbors} $u$, regardless of whether their $h^{l-1}$ embeddings were updated or not, as long as some of them were updated. \textit{This leads to wasted computation because updating even a single in-neighbor among many requires aggregating the embeddings of all in-neighbors to compute the new embedding for the sink vertex, and this grows exponentially downstream.} 

Next, we describe the incremental computation model of \rpp that avoids this by reducing the number of operations performed during aggregation to be proportional to the number of vertices updated.


\subsection{Incremental Update Propagation in \rpp}\label{subsec:design-inc}
Incremental computation in \rpp follows a similar initial step as RC. It \textit{\textbf{\underline{Ap}plies}} the graph updates that have been received onto the current graph (Fig.~\ref{fig:arch}, Inset), and identifies the root vertices of these mutations.
We start a BFS traversal from each \textit{root vertex} of an update (\S~\ref{subsec:design-rc}) 
to identify downstream vertices affected at each hop, and whose embeddings need to be incrementally updated. This forms the $L$-hop \textit{propagation tree}; leaves of this tree are the vertices whose predicted labels ($h^3$ embeddings) may be updated.

\subsubsection{Operations on the Propagation Tree}

We perform an iterative vertex-centric computation within the propagation tree, with vertices at distance $l$ from the root vertex participating in \textit{hop} $l$. This is similar to a \textit{Bulk Synchronous Parallel (BSP)} model of execution~\cite{pregel,giraph}, with $L$ iterative supersteps (hops) of execution.
We employ a strictly \textit{look-forward} computing model to propagate updates from a vertex through \textit{message-passing}, with barrier-synchronization between hops.

In a superstep $l$, we perform several operations on each vertex at hop $l$ as part of its lifecycle: \textit{\textbf{\underline{C}ompute}}, \textit{\textbf{\underline{P}repare}}, \textit{\textbf{\underline{S}end}} and \textit{\textbf{\underline{Ag}gregate}} (Fig.~\ref{fig:arch}, Inset).
The vertex \textit{computes} its updated embeddings~($h^{l}$) based on its current embeddings~($h^{l-}$) and the incoming messages from the previous hop, for $l > 0$. It then \textit{prepares} the outgoing messages to be sent to its out-neighbors, and then \textit{sends} these messages to the sink vertices. These messages carry incremental updates rather than the raw embeddings, as we describe later. This repeats until $L$ hops are complete. 
Unlike sampling-based methods, all vertices in the propagation tree are updated to ensure deterministic behavior.

\subsubsection{Inboxes in \rpp}
Each vertex maintains $L$ logical \textit{inboxes}, one for each hop at which it may be present from a root vertex being updated. Affected in-neighbors at hop $(l-1)$ of the propagation tree will place their incremental messages to this vertex in its hop-$l$ inbox. The inbox will \textit{aggregate} this message with other messages from other in-neighbors in this hop, and maintain just one aggregate message.
E.g., in Fig.~\ref{fig:ripple-single-schematic}(c) \textit{A} has $3$ inboxes that are used to receive messages for its $h^1$ (from $h^0$ of $C$), $h^2$ (from $h^1$ of $C$) and $h^3$ (from $h^2$ of $B,C,D$), based on the level in the tree where $A$ appears.

Messages in the inbox at hop $l$ are processed by vertices after all vertices in the previous hop $(l-1)$ have completed their compute and send operations, consistent with the barrier in BSP~\cite{pregel, yuan2024comprehensive}, which alternates between compute and communication phases.
Such a ``push'' based approach, where only upstream vertices that are affected send messages to the inbox, avoids each vertex checking \textit{all} its in-neighbors for any changes, as is done in a layer-wise recomputation approach.

\subsubsection{Incremental Computation}
An incremental message sent from a source vertex at hop $l$ to its sink vertices at $(l+1)$ is meant to \textit{nullify} the impact of the old embeddings~($h^{l-}$) of the source vertex and \textit{include} the contribution of its current embeddings~($h^{l}$) on the sink vertices, thus forming a ``delta''. 
Propagation of updates in \rpp can happen due to different update conditions. We describe these, and their consequent actions, next using a $sum$ aggregation function as an example.

\paragraph{The embedding of vertex $u$ is updated from $h^{l-}_u$ to $h^{l}_u$} 
This can occur due to a direct update to the feature of vertex $u$~($h^0_u$ gets updated), or because $u$ falls at hop $l < L$ of the propagation tree of another edge or vertex update. If $u$ is not at the leaf of the tree,
it will send a message $m^{l + 1}_{uv}$ to the ($l+1$) inbox of its sink vertex $v$.
The contents of $m^{l + 1}_{uv}$ are \textit{prepared} such that they invalidate the effect of the old embedding $h^{l-}_u$ on $h^{l+1}_v$ and include the contribution from its new embedding $h^{l}_u$.
For the \textit{sum} operator this incremental message is $m^{l + 1}_{uv} = h^{l}_u - h^{l-}_u$.

\begin{figure}[t]
    \centering
    \includegraphics[width=0.8\columnwidth]{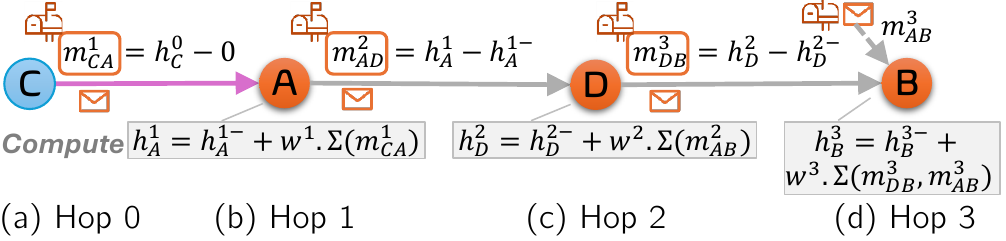}
    \caption{Messages ($m^{hop}_{edge}$) propagate from edge addition $(C,A)$ to update the $h^3$ embedding of vertex $B$~(Fig.~\ref{fig:ripple-single-schematic}), for a 3-layer GNN using \textit{sum}. They negate the effect of the old embeddings ($h^{1-}_A,h^{2-}_D, h^{3-}_B$) to compute the new ones ($h^{1}_A,h^{2}_D, h^{3}_B$).}
    \vspace{-0.1in}
    \label{fig:incremental-update-example}
\end{figure}

E.g., in Fig.~\ref{fig:incremental-update-example}(c), vertex $D$ receives the incremental message $m^2_{AD} = (h^1_A - h^{1-}_A)$ from $A$. When the update is applied to $h^{2-}_D$, it computes the new embedding by cancelling the terms, $h^2_D = h^{2-}_D + w^2\cdot m^2_{DA} = h^{2-}_D + w^2\cdot (h^1_A - h^{1-}_A) = w^2 \cdot (h^1_C + h^{1}_A)$, where $h^{2-}_D= w^2\cdot (h^1_C + h^{1-}_A)$). This value of $h^2_D$ computed by our incremental approach with \textit{$1$ subtraction and $1$ addition} is exactly what would have been computed by layer-wise recomputation, but using \textit{$k$ additions} over the embeddings from $k$ in-neighbors.

\paragraph{An edge $(u, v)$ was added/deleted to/from the graph} Like the previous case, here again $m^{l + 1}_{uv}$ will nullify the effect of $h^{l-}_u$ on $h^{l+1}_v$. But before the edge was added, the embedding $h^{l+1}_v$ did not have any contribution from vertex $u$. So, this is a simplified variant of the previous case where we use a value of $h^{l-}_u=0$ as the old embedding of $u$ when preparing the incremental message~(e.g., $h^{0-}_C$ is used as 0 for $m^1_{CA}$ in Fig.~\ref{fig:incremental-update-example}(a)). Similarly, deleting an edge $(u, v)$ means that $v$ must not get any contribution from $u$, and hence we use $h^l_u=0$ as the new embedding of $u$ within $m^{l + 1}_{uv}$.

\paragraph{Aggregation of message at the inbox} Lastly, a vertex $v$ can receive messages from multiple vertices at hop $l$. Due to the inherent \textit{permutation-invariance} of the GNN aggregation functions due to their commutative property, the arriving messages can be aggregated in a vertex's inbox in any order. For the \textit{sum} operator, this aggregation is simply the sum of all incoming messages.
In Fig.~\ref{fig:incremental-update-example}(d), besides $m^3_{DB}$, vertex $B$ also receives a message $m^3_{AB}$ from $A$ due to an update of $h^{2-}_A$ to $h^2_A$. The messages that arrive at $B$'s hop-$3$ inbox are \textit{added} as they are received. The resultant $h^3_D$ is identical to the outcome if the updates had been applied individually.

\paragraph{Other linear aggregators} This incremental computation model can easily be generalized to other linear aggregators beyond $sum$. For both $mean$ and $weighted~sum$ each message is prepared such that it includes a \textit{weight} $\alpha$ when propagated to a neighbor. The message to propagate a change from $u$ to $v$ at $l$ hop is then modeled as $m^{l + 1}_{uv} = \alpha h^{l}_u - \alpha h^{l-}_u$, where $\alpha$ is in-degree of the sink vertex for mean and the edge-weight for weighted sum, to effectively replace the effect of the old embedding with the new embedding of $u$ at $v$.
Similar to \textit{sum}, message aggregation for linear accumulative messages is simple and just requires adding all incoming messages.

\subsection{Generalization to Complex GNN Architectures}

Till now, we have established the operation of the incremental model for linear aggregators; here, we extend it by identifying scenarios where complex aggregators, such as \textit{min}/\textit{max}, and attention-based~(GAT), require recomputation, and introduce a hybrid incremental programming model used by \rpp to support them.

\subsubsection{Attention-based Architectures}\label{subsubsec:attention}

Graph Attention Networks~(GAT)~\cite{velivckovic2017graph}, which introduces attention mechanisms into graph learning by assigning learnable importance weights to neighbors during aggregation, achieves more powerful learning than GCN. 
The embedding generated by the GAT layer for a vertex $v$ at hop $l+1$ is $h^{l+1}_v = \sum_{u\in\mathcal{N}_{in}(v)}\alpha^{l}_{uv}h^{l}_u$, where:
\begin{align}
\alpha_{uv}^l = \frac{\exp(z_{uv}^l)}{\sum_{u\in\mathcal{N}_{in}(v)} \exp(z_{uv}^l)},~
z_{uv}^l &= a^l \cdot (W^l h_{u}^l \,\|\, W^l h_{v}^l)
\label{eqn:gat}
\end{align}
and $a^l$ and $W^l$ are learnable model parameters.
Unlike other GNNs, GAT's hop-$l$ coefficient for an edge $(u,v)$, $\alpha^l_{uv}$, depends jointly on the $h^l$ embeddings of both source and sink vertices~(Eqn.~\ref{eqn:gat}). Also, changes to $h^l_v$ will trigger changes to downstream embedding, $h^{l+1}_v$, and cause
$z_{uv}^l$ to be calculated for all $u \in \mathcal{N}_{in}(v)$, inevitably requiring access to all in-neighbors, which reduces to \textit{recompute}. 
But, for out-neighbors of $v$ at hop-($l+1$) that have \textit{not yet been updated}, the embeddings can be updated \textit{incrementally}.

\begin{figure}[t]
    \centering
    \includegraphics[width=0.65\columnwidth]{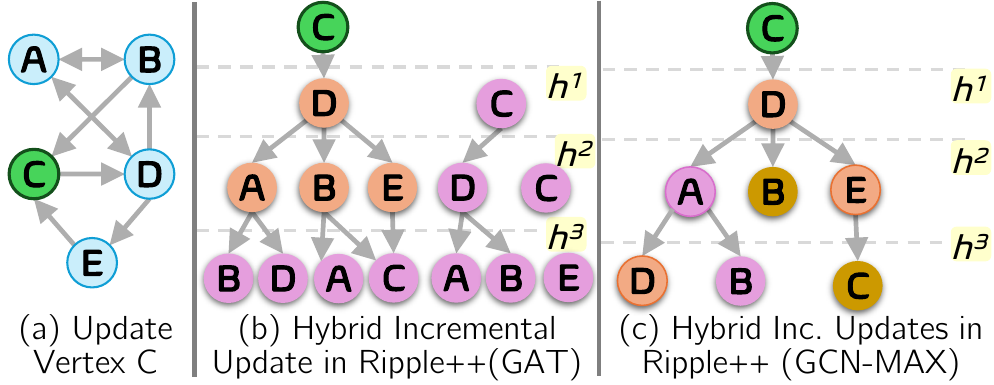}
    \caption{Propagation of updates in the \textit{hybrid-incremental} model of \rpp. Orange vertices are updated incrementally, purple are recomputed, and gold are unchanged for \textit{max} operator.}
    \vspace{-0.1in}
    \label{fig:hybrid-incremental}
\end{figure}

In Fig.~\ref{fig:hybrid-incremental}, the GAT embedding for layer $(l+1)$ of vertex $D$ is $h^{l+1}_D = \big(\alpha_{AD}^lh^l_A + \alpha_{CD}^lh^l_C\big)$. The coefficients $\alpha_{uD}^l$, for all $u \in \{A, C\}$, are computed using Eqn.~\ref{eqn:gat}.
When simplified, $h^{l+1}_D = \frac{\sum_u\exp(z_{uD}^l) h^l_u}{\sum_{u} \exp(z_{uD}^l)}$. Due to an update from $h^{0-}_C$ to $h^0_C$, the only parts of $h^{1}_D$ affected are $\exp(z^0_{CD})h^0_C$ in the numerator and $\exp(z^0_{CD})$ in the denominator.
So, the incremental message sent to any vertex will have two parts: numerator and denominator. E.g., $C$ prepares the message: $m^1_{CD} = \big\langle (\exp(z^{0-}_{CD}) h^{0-}_C - \exp(z^{0}_{CD})h^{0}_C),~(\exp(z^{0-}_{CD}) - \exp(z^{0}_{CD})) \big\rangle$.

\begin{figure}[t]
    \centering
    \includegraphics[width=0.68\columnwidth]{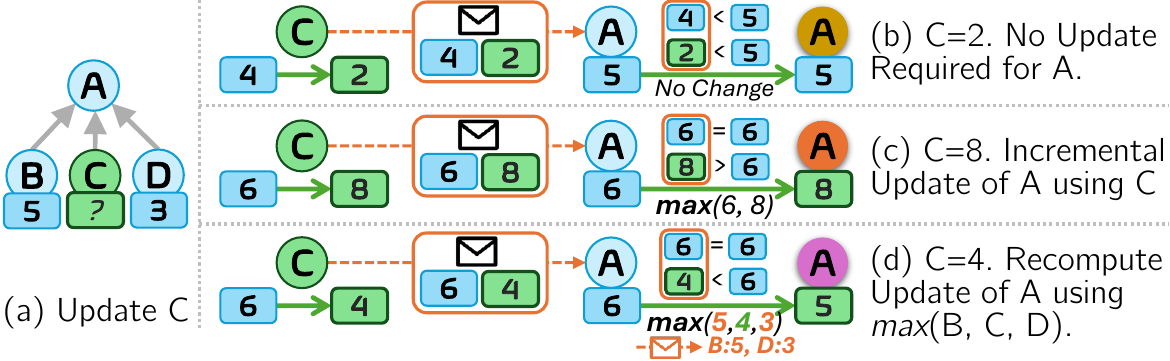}
    \caption{Propagating embeddings for a monotonic \textit{max} agg.}
    \vspace{-0.1in}
    \label{fig:max-cases}
\end{figure}

We propose a \textit{hybrid-incremental} approach for GAT.
Any vertex $u$ that appears at hop $l$ of the propagation tree performs \textit{recomputation} from hop $(l+1)$ onward~(Fig~\ref{fig:hybrid-incremental}(b), \textit{purple} vertices). For vertices that have not been updated yet in the tree, we use \textit{incremental}~(\textit{orange}).

\subsubsection{Monotonic Aggregation Functions}\label{subsubsec:monotonic}
Monotonic aggregators like \textit{max} and \textit{min}, unlike their accumulative counterparts, cannot be updated through local deltas since the effect of an update is apparent only when it reaches the current hop~(Fig.~\ref{fig:max-cases}).
They require recomputation if an update invalidates the current extrema's contribution, but can be done incrementally if the update retains it, as we discuss next.

\rpp extends the above hybrid-incremental approach to GNNs with monotonic aggregation functions. A message sent from vertex $u$ to $v$~($m^{l+1}_{uv}$) at hop $l+1$ contains both the old and the new $l$-hop embedding of the vertex $u$, $h^{l-}_u$ and $h^l_u$.
When vertex $v$ receives a message $m^{l+1}_{uv} = \langle h^{l-}_u, h^l_u\rangle $, it can take one of three actions, illustrated in Fig.~\ref{fig:max-cases} using \textit{max}.
\paragraph{No update is computed for $v$} \label{subsubsec:monotonic:1}
In the trivial case, where the new embedding $h_u^l$ is identical to the current embedding of $v$, an update is not needed. Importantly, if the old embedding $h_u^{l-}$ did not contribute to the current embedding of $v$, and the new embedding $h_u^l$ does not contribute either, the update from $u$ is irrelevant to $v$ and no change is performed at $v$. No further propagation is required. E.g., in Fig~\ref{fig:max-cases}(b), the current embedding of $A$ did not have any contribution from the old embedding of $C$~(value $4$) before the update, and will not be impacted by the update to $C$~(value $2$) either. Here, the embedding at $A$ remains unchanged at $5$ (gold color), and $A$ does not propagate changes further.
\paragraph{Incremental update is computed at $v$}\label{subsubsec:monotonic:2}
The new embedding $h_u^l$ is greater than the current embedding at $v$. In this case, $v$ incrementally updates its embedding by replacing the old contribution from $u$ with the new one, avoiding a full recomputation. E.g., in Fig.~\ref{fig:max-cases}(c), the current embedding of $A$ included a contribution from the old embedding of $C$~(value $6$), and the updated value from $C$~(value $8$) can cover the previous value. Hence, the new embedding of $A$ (e.g., $8$) can be computed incrementally.
\paragraph{Recompute performed at $v$}\label{subsubsec:monotonic:3}
The old embedding $h^{l-}_u$, contributed to the current embedding of $v$, but the new embedding $h^l_u$
no longer preserves the monotonic consistency required for performing an incremental update,
e.g., with a \texttt{max} aggregator, the new value may no longer be the maximum among all neighbors. Here, $v$ must recompute its embedding from all current contributors to ensure correctness. In Fig.~\ref{fig:max-cases}(d), the old embedding of $C$~(value $6$) was the highest among the neighbors of $A$, but the updated embedding~(value $4$) is no longer the maximum. Hence, $A$ recomputes its embedding by taking the new maximum from among all its neighbors~($5$, from $B$), resulting in its new value of $5$. This change is propagated as needed.

\rpp 
splits the active vertices at each hop into three sets matching these scenarios for processing -- vertices that do not need updates~(\textit{gold} in Fig.~\ref{fig:hybrid-incremental}(c)) are left unchanged and do not propagate 
further; vertices that need recomputation~(\textit{purple}) pull the embeddings of all their in-neighbors; and vertices that are updated incrementally~(\textit{orange}) apply the message to get the new embedding.

\begin{figure}[t]
    \centering
    \subfloat[Recompute (RC)]{\includegraphics[width=0.8\linewidth]{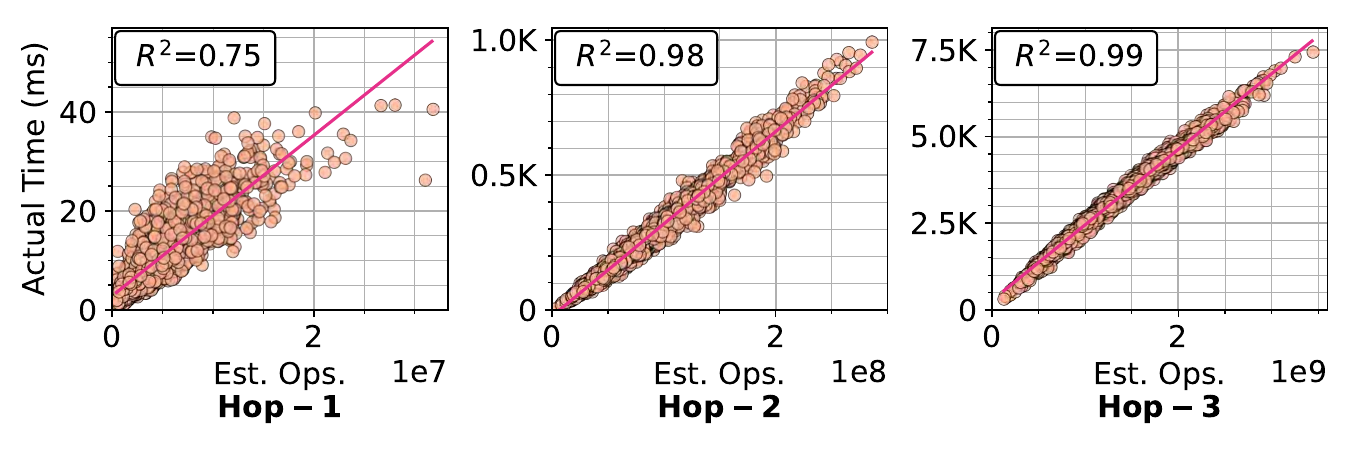}\label{subfig:benefits-rc}}\\
    \vspace{-0.1in}
    \subfloat[\rpp (RP)]{\includegraphics[width=0.8\linewidth]{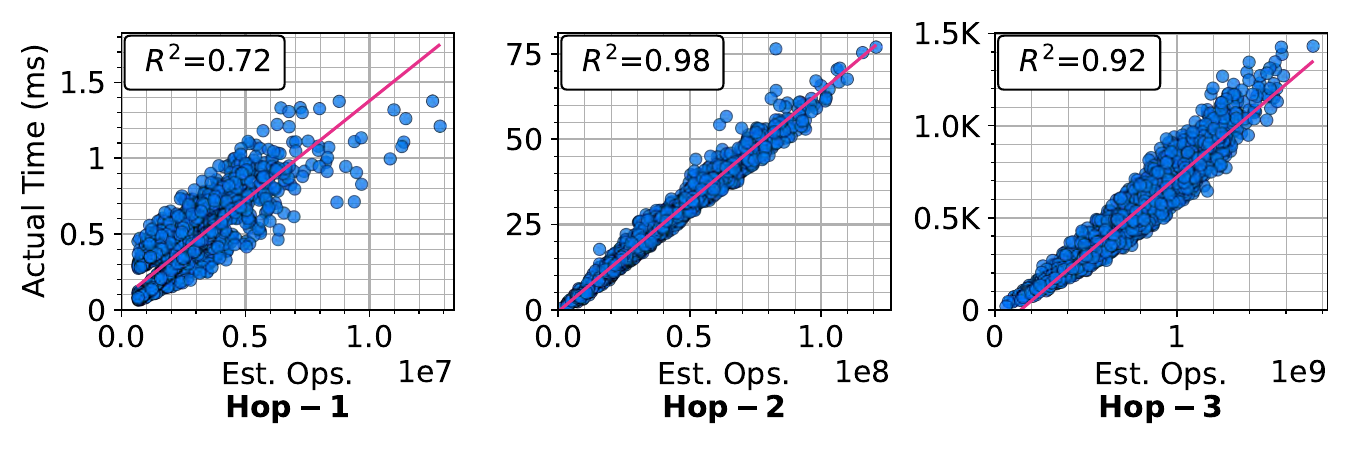}\label{subfig:benefits-rp}}
    \caption{Benefits analysis of \rpp over \rc for inferencing a 3-layer GC-S for Products graph with $bs=100$.}
    \vspace{-0.1in}
    \label{fig:benefits-analysis}
\end{figure}

\subsection{Analytical Modeling of \rpp Benefits over RC}\label{subsubsec:benefits-analysis}
Next, we analytically examine the advantages of \rpp over layer-wise recomputation~(\textbf{RC}). 
Let $a_l$ denote the number of active vertices at hop-$l$ of the propagation tree, i.e., vertices whose $h^l$ embeddings require updating. Let $\delta$ be the average degree of the graph and $d_l$ be the dimensionality of the embedding at hop-$l$. For simplicity, we assume a \textit{sum} aggregation function in this analysis.

In RC, updating the embeddings of $a_l$ active vertices at hop-$l$ involves two primary steps. 
\begin{enumerate}[leftmargin=*]
    \item \textit{Aggregate in-neighbors:} Here, the $h^{l-1}$ embeddings of the in-neighbors of each active vertex are aggregated, requiring $(a_l \cdot \delta \cdot d_{l-1})$ floating-point (FP) operations.
    \item \textit{Compute:} Here, the aggregated embeddings of size $a_l \times d_{l-1}$ are multiplied with the weight matrix of size $d_{l-1} \times d_l$, incurring $(a_l \cdot d_{l-1} \cdot d_l)$ FP operations.
\end{enumerate}
Thus, the total estimated cost of RC to process vertices at hop-$l$ is
$
\big(a_l \cdot \delta \cdot d_{l-1} + a_l \cdot d_{l-1} \cdot d_l\big)
$.

Conversely, in \rpp, updating the embeddings of $a_l$ active vertices at hop-$l$ involves two primary steps. 
\begin{enumerate}[leftmargin=*]
    \item \textit{Prepare and aggregate message:} Here, messages are generated for the active vertices in hop $(l-1)$, with each of the $a_{l-1}$ active vertices sending a message to its out-neighbors and they aggregating it in their inbox. This requires
$\big( a_{l-1} \cdot d_{l-1} + a_{l-1} \cdot \delta \cdot d_{l-1} \big)$ 
FP operations.
    \item \textit{Compute:} Here, the aggregate messages are used to compute the embeddings for hop-$l$ by multiplying them by the weight matrix and adding the result to the current embedding, costing $a_l \cdot d_{l-1} \cdot d_l + a_l \cdot d_l$ FP operations.
\end{enumerate}
The total estimated compute cost for \rpp to process $a_l$ vertices at hop-$l$ is
$
\big(a_{l-1} \cdot \delta \cdot d_{l-1} + a_{l-1} \cdot d_{l-1} + a_l \cdot d_{l-1} \cdot d_l + a_l \cdot d_l\big)
$,
which can further be approximated down to
$
\big(a_{l-1} \cdot \delta \cdot d_{l-1} + a_l \cdot d_{l-1} \cdot d_l\big)
$.

Thus, RC requires 
$\approx(a_l \cdot \delta \cdot d_{l-1} + a_l\cdot d_{l-1}\cdot d_l)$ FP operations at hop~\(l\), whereas \rpp requires $\approx(a_{l-1}\cdot \delta\cdot d_{l-1} + a_l\cdot d_{l-1}\cdot d_l)$ FP operations. Since the active vertices can only expand or remain constant ($a_l \ge a_{l-1}$), the difference in cost simplifies to $(a_{l-1} - a_l)\cdot \delta\cdot d_{l-1}$. 
Hence, \rpp is strictly cheaper when the frontier expands~($a_l > a_{l-1}$), and the two methods have identical costs when the frontier size remains unchanged~($a_l = a_{l-1}$). \rpp never exceeds RC's cost.

Fig.~\ref{fig:benefits-analysis} shows a strong correlation between the estimated FP operations and the actual execution time for both RC~(Fig.~\ref{subfig:benefits-rc}) and \rpp~(Fig.~\ref{subfig:benefits-rp}) for GC-S inferencing on the Products graph~(see \S~\ref{subsec:eval-expsetup} for experiment setup). 
The correlation is particularly strong at hop-$2$ and hop-$3$~($R^2 \ge 0.92$), which dominate the total inference time, confirming that our FP count model can reliably predict runtime.
Hop-$1$ exhibits a weaker correlation~($R^2 \approx 0.72$--$0.75$) due to fixed-cost overheads, which are more prominent at small workloads. 
These are, however, very small in absolute times ($<50ms$) compared to 1000s of $ms$ taken by hops $2$ and above.


\section{Distributed GNN Inference in \rpp}
\label{sec:design:distr}

\begin{figure}[t ]
    \centering
    \includegraphics[width=0.99\linewidth]{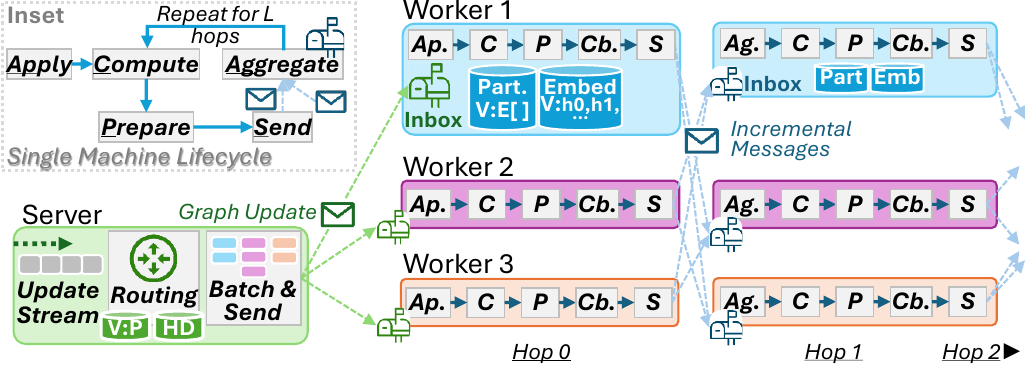}
    \caption{Distributed Architecture of \rpp. Top-left \textit{Inset} shows single machine lifecycle.}
    \vspace{-0.1in}
    \label{fig:arch}
\end{figure}

Real-world graphs can be large in size. Besides the graph structure and features, we also need to store the various embeddings for all vertices to help with the incremental computation. These data structures can be memory-intensive, exceeding the RAM of a single machine, e.g., for Papers, $26$GiB is used for the graph structure ($111\M$ vertices, $1.6\B$ edges), $56$GiB for features, and $132$GiB for intermediate embeddings created when computing a 3-layer GNN ($>200$GiB in total). We design a distributed execution model for \rpp (Fig.~\ref{fig:arch}) to perform incremental computation over updates received for the graph partitioned across machines in a cluster, while retaining an execution flow similar to a single-machine (Fig.~\ref{fig:ripple-single-schematic}).

\subsection{Partitioning the Graph}\label{subsubsec:dist-partition}
We consider a typical scenario where the initial part of the graph has arrived, and the updates start streaming in. This initial graph structure, features, and their embeddings computed for the GNN are partitioned and kept in-memory on \textit{worker} machines~\cite{zheng2020distdgl, cai2021dgcl, ma2019neugraph, zhang20212pgraph, ramezani2022learn}. We use METIS~\cite{karypis1997metis} to partition the initial graph so that the vertex count is balanced across workers (balances compute load), and edge-cuts across partitions are minimized (reduces network communication), with one partition per worker. We replicate the boundary (``halo'') vertices having edge cuts to reduce communication during BFS propagation~\cite{zheng2020distdgl}.

\subsection{Server-Worker Execution Model}\label{subsubsec:dist-exec-model}
We adopt a worker--server architecture for distributed execution (Fig.~\ref{fig:arch}). The \textit{server} receives the incoming stream of events and routes them to the relevant workers. The server also maintains metadata about the workers, e.g., vertex to partition mapping (V:P), high-degree vertices (HE), for efficient routing.

Each \textit{worker} stores the embeddings for its local vertices, and is responsible for executing the lifecycle for its local vertices (Fig.~\ref{fig:arch}): receive messages, update embeddings, and prepare and send incremental messages. Unlike a single-machine setup where all vertices have inboxes, doing so for large graphs is memory-intensive and unnecessary since only a subset of vertices participate in the propagation tree. Instead, each worker maintains a \textit{managed inbox pool} 
that gets \textit{assigned} to vertices on-demand, when a message is received. After processing a batch, the inboxes are unassigned and can be reassigned to other active vertices in the future. 
We dynamically expand this allocation pool on-demand. This reduces memory overhead from repeated allocation and right-sizes memory use.

Workers do not hold embeddings for \textit{halo vertices}, but maintain inboxes for them. These serve as a proxy for sending messages from local to halo vertices in a partition, and aggregate local partition messages using a MapReduce-style \textit{\textbf{combiner (\underline{Cb.})}} (Fig.~\ref{fig:arch})~\cite{pregel}. After the \textit{local send} operation in a hop, we perform a \textit{remote send} of these aggregate messages to remote workers having their partition. We perform another receiver-side aggregation, and start the next hop's compute.

\subsection{Request Batching and Routing}\label{subsubsec:dist-batching}

The server
receives a stream of both vertex-level and edge-level updates that are routed to appropriate workers. The server creates \textit{batches} of updates destined for each worker, of size of $bs$ (Fig.~\ref{fig:arch}). 
Both \textit{edge updates} are of the form $v_{12}$, which indicates an edge from vertex $v_1$ to vertex $v_2$.
We model vertex \textit{additions} also as an edge update as we assume that at least one edge will be added along with a new vertex.
\textit{Vertex deletions} and \textit{feature updates} apply only to vertices already present in the graph.

A vertex update/delete is easy to route, and is assigned to the worker (partition) having it as a local vertex. The server maintains a \textit{Vertex-Partition map (VP)} from each vertex to its partition.
But edge additions or deletions may span workers.
\begin{itemize}[leftmargin=*]
    \item An edge request is completely local to a worker if both the incident vertices of an edge update are local to the partition. Here, the update is assigned to that worker.
    \item An edge request spans workers if its source and sink vertices are local to different partitions. We assign the edge update to the worker that has the source vertex (or hop-$0$) of the edge update. We also send a \textit{no-compute} request to the sink vertex's worker to update its partition with the new edge.
\end{itemize}

Finally, vertex addition requires more sophisticated handling since it must assign the new vertex to a partition while considering placement quality for load balancing and minimizing edge cuts, similar to graph partitioning. \rpp allows users to configure custom routing algorithms, and we
propose two: (1) A simple \textit{hash-based routing}, where the new vertex is assigned to a partition based on hashing its vertex ID onto a partition ($V\%P$); and (2) A \textit{locality-aware routing}, discussed next, that minimizes edge cuts while balancing load across partitions.
Once $bs$ streaming updates for a batch have arrived in total,
the server sends them to the respective workers.

\begin{algorithm}[t]
\caption{Locality-aware Routing and Vertex Placement}
\label{algo:placement}
\begin{algorithmic}[1]
\Procedure{AssignEdge}{Edge $v_{12}$, Partitions $P_i$} 
\State $\pi(v)$: Partition assigned to vertex $v$
\State $\text{load}(P_i)$: Number of vertices assigned to partition $P_i$
\State $H(v)$: Lookup function, returns True if $v$ is a \textit{high in-degree} vertex, False otherwise
\If{$v_1,v_2$ exist} \label{algo:placement:v1v2exist}
   \If{$\pi(v_1)=\pi(v_2)$} $\pi(v_{12}) \gets \pi(v_1)$ \label{algo:placement:v1v2same}
   \Else{} \Call{CutEdge}{$\pi(v_1),\pi(v_2)$} \label{algo:placement:v1v2diff}
   \EndIf
\ElsIf{$v_2$ exists} \label{algo:placement:v2exists}
   \If{$H(v_2)$} $\pi(v_{12}) \gets \pi(v_2)$ \label{algo:placement:v2high}
   \Else{} $\pi(v_{12}) \gets \arg\min_i \{\text{load}(P_i)\}$ \label{algo:placement:v2balance}
   \EndIf
\Else{} $\pi(v_{12}) \gets \arg\min_i \{\text{load}(P_i)\}$ \label{algo:placement:balance}
\EndIf
\EndProcedure
\end{algorithmic}
\end{algorithm}

\subsection{Locality-aware Routing for Updates}\label{subsubsec:dist-placement}
We design a locality-aware routing strategy for edge-events, inspired by our prior work~\cite{bhoot2025triparts}. Alg.~\ref{algo:placement} running on the server decides which partition to place an incoming edge $v_{12}$, caused by an edge addition/deletion or vertex addition as part of a new edge.
\begin{itemize}[leftmargin=*]
    \item If both incident vertices $v_1$ and $v_2$ already exist in partitions, this is either an edge addition or deletion~(line~\ref{algo:placement:v1v2exist}). Here, if $v_1$ and $v_2$ are on the same partition~(line~\ref{algo:placement:v1v2same}), edge $v_{12}$ is sent to that partition, avoiding inter-worker communication. 
    \item If the incident vertices are on different partitions, a cut-edge is introduced between partitions $\pi(v_1)$ and $\pi(v_2)$~(line~\ref{algo:placement:v1v2diff}).
    This routes the edge to the source vertex partition but sends a copy to the other partition to update the graph.
    \item If only one vertex exists, $v_2$, then the other vertex $v_1$ is new, and indicates a vertex addition~(line~\ref{algo:placement:v2exists}). We consider if the incident vertices of $v_{12}$
    have a \textit{high in-degree} -- co-locating the edge with the vertex with a high degree enhances locality and reduces the inter-worker communication for the propagation tree. We maintain a \textit{High in-Degree set (HD)} at the server for this.
    If the existing vertex $v_2$ is part of the HD set, $v_{12}$ is placed on its partition (line~\ref{algo:placement:v2high}). Else, we place the edge on the partition with the least load in an effort to \textit{balance} the vertex load across partitions~(line~\ref{algo:placement:v2balance}).
    \item If neither vertex exists~(line~\ref{algo:placement:v2balance}), the edge is assigned to the least-loaded partition, again to balance the vertex load.%
\end{itemize}%
\noindent The HD set (Fig.~\ref{fig:arch}) has vertices with degree greater \textit{than twice the current average degree} of the graph, and is refreshed to reflect structural changes, e.g., after each $0.5\M$ updates.
Hash-based routing has constant \textit{time and space complexity}. Locality-based routing also has constant-time lookups, but incurs $\mathcal{O}(V)$ time to refresh the HD set. It also uses $\mathcal{O}(V)$ space to track vertex membership and HD nodes.
Overall, this strategy ensures that we systematically minimize communication, maintain load balancing, while ensuring rapid routing of streaming updates in large distributed graphs.

\subsection{Update Processing}\label{subsubsec:dist-processing}
Upon receiving a batch of updates, a worker applies the topology and feature changes to its local partition and embeddings~(\textit{\underline{Ap}ply} in Fig.~\ref{fig:arch}). 
Depending on the update, it prepares messages for the hop-$1$ vertices (\textit{\underline{Pr}epare}.
For local vertices, the message handling is similar to a single-machine execution. But for halo vertices, the messages are \textit{combined (\underline{Cb.})} in a local mailbox for the halo, across all propagation trees at hop-$0$. Once all the vertices in this hop are processed by the worker, the combined messages are sent to matching vertices on remote workers, before the next hop starts.

For subsequent hops, $l \in [1, L]$, the workers aggregate the messages in the vertex inboxes, compute the local embeddings, and prepare downstream messages, similar to a single machine; as before, local messages are delivered and aggregated immediately, while remote messages are combined and sent to the remote inbox after the hop completes. We leverage MPI for efficient message passing between workers.
The next hop starts only after embeddings in the prior hop are updated, and incremental messages are delivered to all workers. This repeats $L$ times till we get the final layer embeddings for 
the batch.


\section{\rpp Implementation and Baselines}\label{sec:implementation}
\rpp is implemented natively in Python for both single-machine and distributed setups using NumPy v$2.0$, and is open-sourced \footnote{\url{https://github.com/dream-lab/ripple}}. The distributed setup utilizes MPI for inter-process communication. We only support CPU-based execution due to its competitive performance relative to GPU, as discussed later. We also implement the \textit{vertex-wise} and \textit{layer-wise recompute} inference strategies as baselines using the SOTA DGL v$1.9$ framework with a PyTorch backend~\cite{wang2019deep}. We natively implement the \textit{layer-wise recompute} strategy directly using Python, due to the high overheads for graph updates using DGL. We also compare \rpp against InkStream~\cite{wu2025inkstream}, a SOTA streaming GNN inference framework.

To maintain a fair comparison, all strategies update only the embeddings of impacted vertices at the final hop when performing a batch of updates. We consider the whole neighborhood of a vertex at each hop during inference, without sampling or any approximations, to ensure that the resulting predictions are accurate and deterministic. Finally, we confirm that \rpp calculates \textit{accurate} embeddings at all hops of inferencing, within the limits of floating-point precision.


\section{Experimental Evaluation}\label{sec:evaluation}
We evaluate \rpp on diverse graph datasets and GNN workloads for single-machine and distributed setups, and compare it against contemporary baselines.

\begin{figure*}[t]
    \centering
    \includegraphics[width=\textwidth]{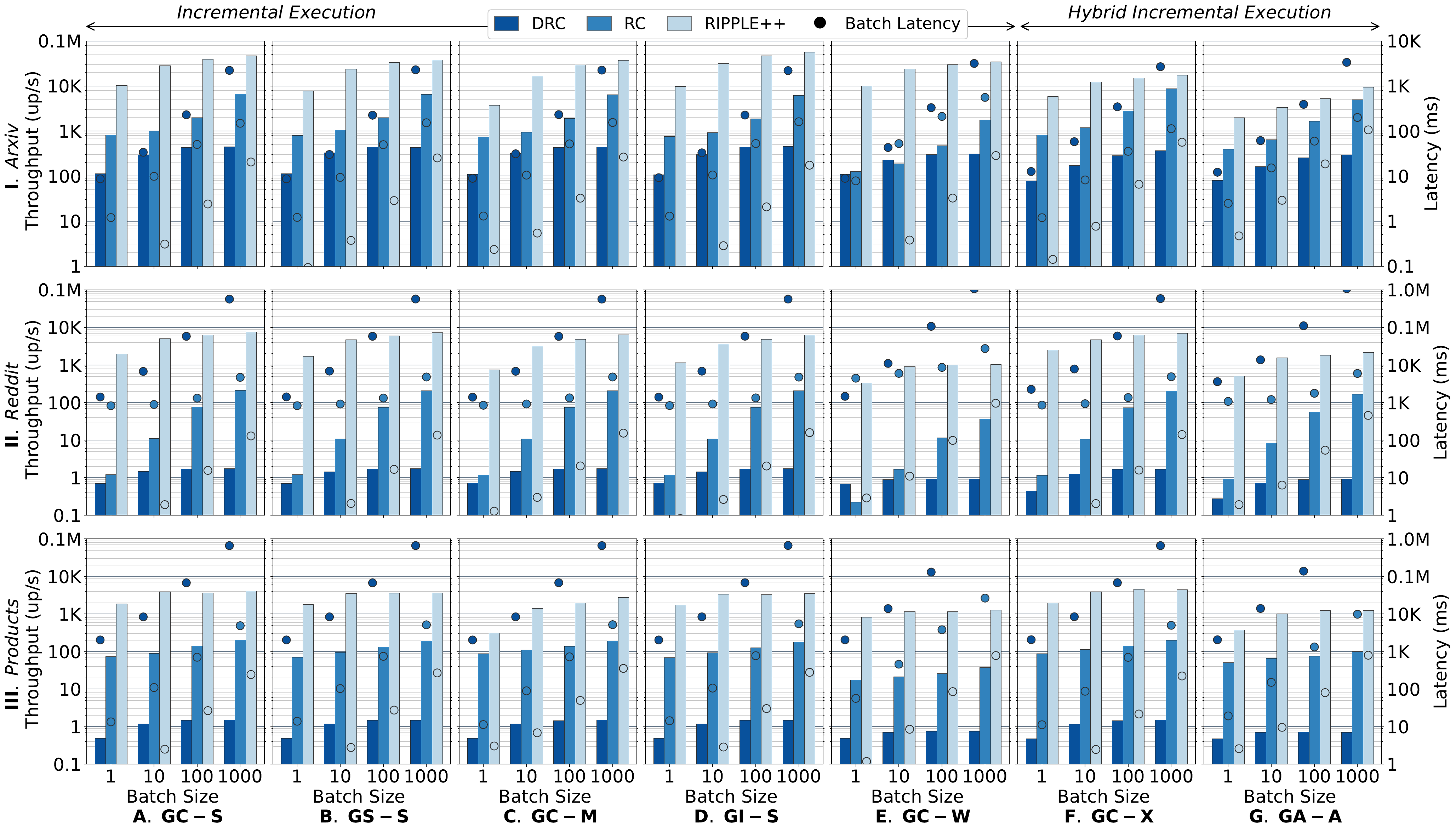}
    \caption{Single machine performance of \rpp against DRC and RC on the 2-layer variants of $7$ GNN workloads for Arxiv, Reddit, and Products across $bs$. Left y-axis shows \textit{Throughput} (updates/second, bar, log scale) while right y-axis has \textit{mean batch latency} (milliseconds, marker, log scale)}
    \vspace{-0.1in}
    \label{fig:results-combined-l2}
\end{figure*}

\subsection{Experimental Setup}\label{subsec:eval-expsetup}

\begin{table}[t]
\centering
\setlength{\tabcolsep}{4pt}
\caption{Graph datasets used in experiments.} 
\begin{tabular}{l||r|r|r|r|r}
\hline
\textbf{Graph} & $|V|$  & $|E|$   & \# Features & \# Classes & Avg. In-Deg.\\ \hline\hline
\textbf{\textit{Arxiv}}~\cite{hu2020open}                & $169\K$   & $1.2\M$  & $128$ & $40$ & $6.9$\\
\textbf{\textit{Reddit}}~\cite{hamilton2017sage}               & $233\K$   &  $114.9\M$  & $602$ & $41$ & $492$\\
\textbf{\textit{Products}}~\cite{hu2020open}         & $2.5\M$   & $123.7\M$ & $100$  & $47$ & $50.5$\\ 
\textbf{\textit{Papers}}~\cite{hu2020open}          & $111\M$   & $1.62\B$ & $128$  & $172$ & $14.5$\\  \hline
\end{tabular}
\label{tab:dataset-specs}
\vspace{-0.1in}
\end{table}

\subsubsection{GNN Workloads}\label{subsubsec:eval-expsetup-workloads}
We evaluate four popular GNN models for vertex classification: \textit{GraphConv}~\cite{kipf2016semisupervised}, \textit{GraphSAGE}~\cite{hamilton2017sage}, \textit{GINConv}~\cite{xu2018powerful}, and \textit{GATConv}~\cite{velivckovic2017graph}. These are paired with common aggregations: linear~(sum, mean, weighted sum), monotonic~(max, min), and attention-based, to construct seven representative workloads:
\underline{G}raph\underline{C}onv with \underline{S}um~(\textbf{GC-S}),
\underline{G}raph\underline{S}AGE with \underline{S}um~(\textbf{GS-S}),
\underline{G}raph\underline{C}onv with \underline{M}ean~(\textbf{GC-M}),
\underline{G}IN\underline{C}onv with \underline{S}um~(\textbf{GI-S}),
\underline{G}raph\underline{C}onv with \underline{W}eighted Sum~(\textbf{GC-W}),
\underline{G}raph\underline{C}onv with Ma\underline{X}~(\textbf{GC-X}), and
\underline{GA}TConv with \underline{A}ttention~(\textbf{GA-A}).

\subsubsection{Datasets}\label{subsubsec: eval-datasets}
We use $4$ well-known graph datasets for our evaluation (Table~\ref{tab:dataset-specs}): ogbn-arxiv~(\textit{Arxiv}), a citation network~\cite{hu2020open}; \textit{Reddit}, a social network~\cite{hamilton2017sage}; ogbn-products~(\textit{Products}), an e-commerce network~\cite{hu2020open}; and a large citation network, ogbn-papers100M~(\textit{Papers})~\cite{hu2020open}, which does not fit in the memory of a single machine and is used only for distributed execution.
We remove a random $20\%$ of vertices and edges from each graph, and the remaining $80\%$ of vertices and edges serve as the initial graph snapshot to which streaming updates arrive. 
We train the GNN models on these $80\%$ snapshots and extract the embeddings, which form the initial state for inference.
In the distributed setup, we partition the graph using METIS~\cite{karypis1997metis} into the required number of parts and load the local subgraph, their embeddings, and the halo vertices in memory.

We generate a stream of events comprising the five types of updates using the probability distribution of these update types from the Facebook workload~\footnote{\url{https://github.com/facebookarchive/linkbench/blob/master/config/FBWorkload.properties}}.
For the single-machine experiments, we generate $200\K$ update events for Arxiv, and $20\M$ events each for Products and Reddit, while for distributed experiments on Papers, we generate $5\M$ events. 
Importantly, given the Facebook distribution of update types, with additions exceeding deletions, the graph size never decreases as events are applied; it either grows or remains unchanged.

\subsubsection{Hardware Setup}\label{subsubsec: eval-testbed}

All single-machine experiments are conducted on workstations with $12$-core AMD Ryzen 9 7900X processor~($4.7$GHz) with $128$~GiB of RAM. 
The distributed execution is performed on a cluster of $24$ compute servers, each equipped with a $16$-core Intel Xeon Gold CPU~($2.9$GHz) with $128$~GiB of RAM, and connected over 10~Gbps Ethernet.
We also briefly compare the DGL implementation of vertex-wise and layer-wise recompute using both CPU and GPU to demonstrate that the GPU-based execution is occasionally slower than the CPU-based execution, and that both DGL variants are slower than our custom implementation of layer-wise recompute on the CPU. For this, we use the same workstation as above with an NVIDIA RTX $4090$ GPU card with $24$~GiB GPU memory.

To ensure fair comparison, we benchmark our implementations of DGL's vertex-wise inference on the CPU~(\dnc) and CPU+GPU~(\dng), with its layer-wise recompute strategy on the CPU~(\drc) and CPU+GPU~(\drg). We show that \drc and \rc running on the CPU act as the most competitive baselines to compare against \rpp's incremental approach. For brevity, we report this experimental comparison 
in the Appendix~\ref{appendix:A}.

\begin{figure*}[t]
    \centering
    \includegraphics[width=\textwidth]{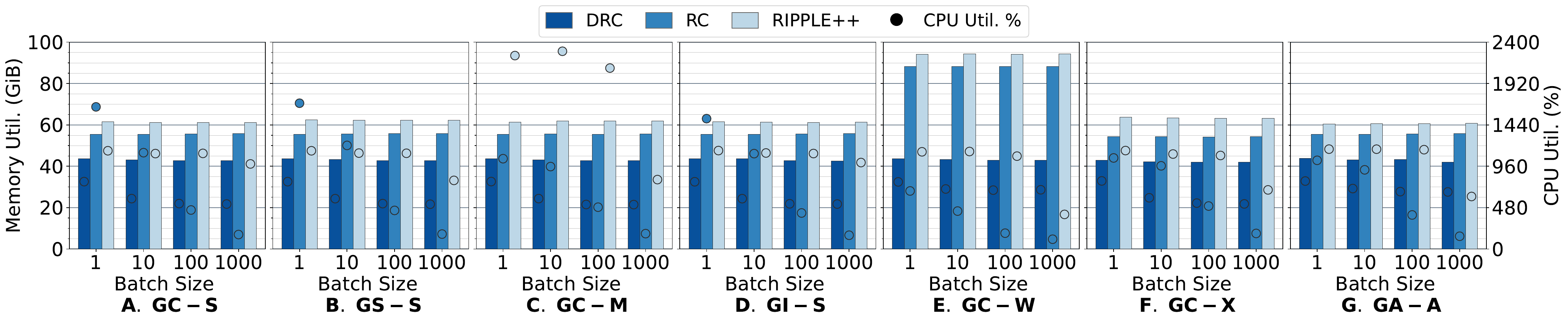}
    \caption{Single-machine systems utilization of \rpp against DRC and RC on the 2-layer variants of $7$ GNN workloads on Products. Left y-axis shows \textit{Memory Utilization}~(GiB, bar) while right y-axis has \textit{CPU Utilization}~(\%, marker)}
    \vspace{-0.1in}
    \label{fig:results-combined-l2-util}
\end{figure*}

\subsection{Single-Machine Performance}\label{subsec:eval-single}
\subsubsection{Comparison With Baselines}\label{subsec:eval-single-baselines}
We contrast the single-machine performance of \rpp~(RP)
with the DRC and RC baselines,
for Arxiv, Reddit, and Products graphs, for the $7$ GNN workloads using different $bs$. All experiments are run for a maximum of $4$~hours or till all events are exhausted.
Fig.~\ref{fig:results-combined-l2} shows the 
\textit{throughput} (\textit{bar, left y-axis}) and \textit{batch latency} (\textit{marker, right y-axis}), for the three graphs shown in three rows, and similarly Fig.~\ref{fig:results-combined-l2-util} plots their \textit{memory} (\textit{bar, left y-axis}) and \textit{CPU utilization} (\textit{marker, right y-axis}).
For the first $5$ workloads~(GC-S, GS-S, GC-M, GI-S, and GC-W)
in Figs.~\ref{fig:results-combined-l2} and~\ref{fig:results-combined-l2-util}~(A--E), 
\rpp performs \textit{purely incremental} processing, while for GC-X and GA-A~(F and G), a \textit{hybrid incremental} approach is required.

\paragraph{Arxiv}
\rpp~(RP) achieves a maximum speedup of $\approx128\times$ over RC and $\approx124\times$ over DRC~(Fig.~\ref{fig:results-combined-l2}, first row). The maximum absolute throughput achieved by RP exceeds $56\K$~up/s for $bs=1000$ for GI-S~(\textit{bar, left y-axis}) with latencies between $0.06$--$104$~ms per batch~(\textit{bar, left y-axis}), across all $bs$ and GNNs. As expected, the absolute throughput values averaged across all $7$ workloads increase from $\approx8\K$~up/s to $\approx35\K$~up/s as $bs$ increases from $1$ to $1000$, since larger batches allow for more efficient amortization of per-batch overheads and hence higher throughput. However, this comes at the cost of higher batch latencies. This offers an opportunity to trade-off throughput and latency according to application requirements.

\paragraph{Reddit}
For Reddit~(Fig.~\ref{fig:results-combined-l2}, second row), increasing $bs$ from $1$ to $1000$ leads to an average throughput increase across all workloads, from about $\approx1.3\K$~up/s to $\approx5.5\K$~up/s. 
RP achieves a maximum speedup for GC-S of $1628\times$ over RC for $bs=1$ and $4393\times$ over DRC for $bs=1000$, with a peak of $\approx7.8\K$~up/s seen for GC-S at $bs=1000$. 
Batch latencies range from $0.3$--$960$~ms across all $bs$ and GNNs.

\paragraph{Products}
For Products~(Fig.~\ref{fig:results-combined-l2}, third row), RP achieves a peak speedup for workload GC-S of $\approx45\times$ over RC for $bs=10$ and $3800\times$ over DRC for $bs=1$. 
The maximum throughput reaches $\approx4.5\K$~up/s at $bs=1000$ for workload GC-X, with latencies of $0.5$--$800$~ms.
On average, throughput across all seven GNNs increases from roughly $1.3\K$~up/s to $\approx3\K$~up/s as the $bs$ grows from $1$ to $1000$, with a corresponding rise in batch latencies.

Overall, \rpp offers throughput up to $56\K$~up/s with latencies of $0.06$--$960$~ms, that translate to a throughput speedup $20$--$438\times$ over RC and $75$--$2771\times$ over DRC, averaged over datasets, GNNs, and batch sizes, thus making it compelling for real-time GNN inferencing applications.

\begin{figure*}[t]
  \centering
  \begin{minipage}[t]{0.68\textwidth}
    \centering
    \includegraphics[width=\textwidth]{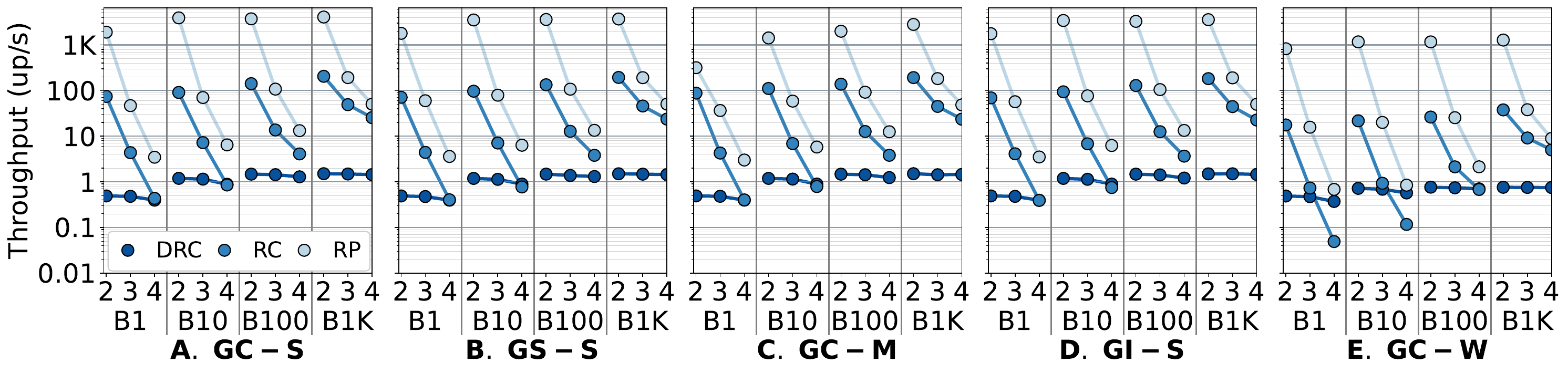}
    \caption{Single machine \textit{Throughput} (log scale) of \rpp (RP) against DRC and RC on the 2-, 3-, and 4-layer variants of $5$ GNN workloads for \textit{Products}, across $bs$.}
    \label{fig:results-products-l2-l3-l4}
  \end{minipage}~~
  \begin{minipage}[t]{0.30\textwidth}
    \centering
    \includegraphics[width=\textwidth]{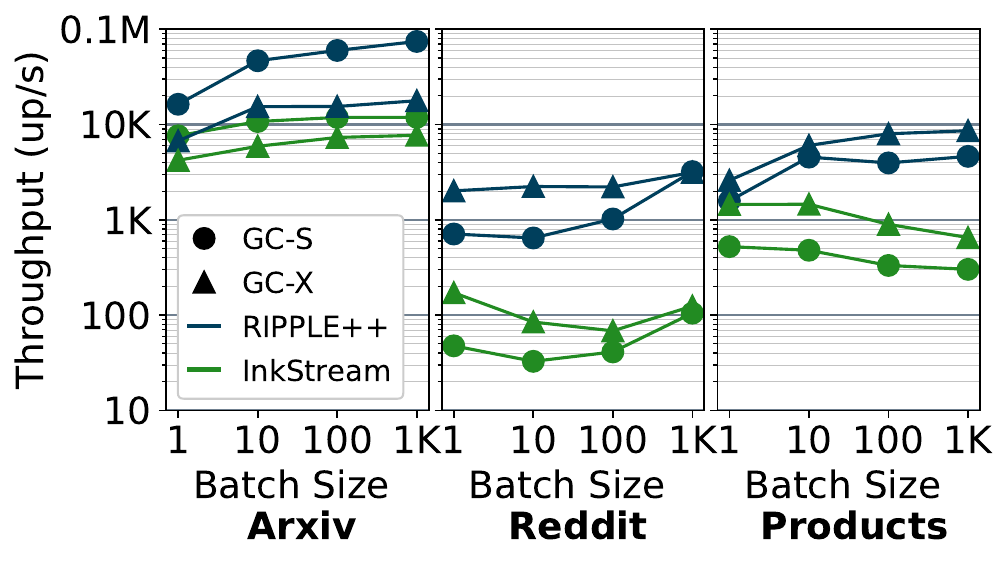}
    \caption{Throughput of \rpp and InkStream for $2$-layer GC-S and GC-X.}
    \label{fig:ink-v-ripple}
  \end{minipage}
  \vspace{-0.1in}
\end{figure*}

\subsubsection{Performance Drill-down on Single-Machine}
Next, we dive deeper into the performance of \rpp on a single-machine setup and make the following key observations.

\paragraph{Increasing $bs$ gives minimal performance benefits for DRC}
As the $bs$ increases from $1$--$1000$, throughput values increase only by about $\approx3.9\times$ for Arxiv and $\approx2.6\times$ for Reddit and Products on average for DRC, with the increase between $100$ to $1000$ being a meager $8\%$ for Arxiv and $\approx2\%$ for Reddit and Products~(Fig.~\ref{fig:results-combined-l2}). As we see in Fig.~\ref{fig:gpu-analysis}
of Appendix, a significant portion of the total batch processing time is spent applying updates to the DGL graph itself~(\textit{Update} in Fig.~19.
As the $bs$ increases, the overhead of modifying the graph grows proportionally and begins to dominate the end-to-end batch time, thus diminishing the relative benefit of batching for the actual embedding propagation.

\paragraph{GC-W has a lower performance compared to other purely incremental GNN workloads, GC-S, GS-S, GC-M, and GI-S}
GC-W for $bs=1000$ achieves a throughput on RP of $\approx34\K$, $\approx1.1\K$, and $\approx1.3\K$~up/s for Arxiv, Reddit, and Products, respectively, in contrast to $\approx45\K$, $\approx7\K$, $\approx3.5\K$~up/s achieved for GC-S -- GI-S, indicating an average drop of $\approx57.4\%$, from Fig.~\ref{fig:results-combined-l2}. 
RC also sees a similar average drop of $78\%$ for GC-W, across graphs, and similarly for DRC.
This is due to the additional \textit{weighted aggregation} required in GC-W. Unlike simpler aggregation functions such as \textit{sum} or \textit{mean}, which combine neighbor embeddings directly, weighted aggregation introduces an extra computational step -- each neighbor embedding must first be multiplied by its associated edge-weight before aggregation, significantly increasing the compute overhead. Notably, the weights fetching and multiplication are non-natively vectorized operations, causing a drop in CPU\%~(Fig.~\ref{fig:results-combined-l2-util}, \textit{marker, right y-axis}) for \textit{bs = 1000}, from $\approx880\%$ for the other incremental GNNs to $\approx420\%$ for GC-W; hence, the compute takes longer.

\paragraph{The throughput speedup of RP over RC decreases as the $bs$ increases}
As the $bs$ increases from $10$ to $1000$, the average speedup of RP over RC across all $7$ workloads drops from approximately $35\times$ to only $7\times$ for Arxiv, $388\times$ to $30\times$ for Reddit, and $33\times$ to $20\times$ for Products. As discussed in \S~\ref{subsubsec:benefits-analysis}, the relative advantage of RP over RC diminishes when $a_{l-1}$ approaches $a_l$. This occurs because as $bs$ increases, it causes $a_{l-1}$ to cover a larger part of the graph with greater overlap between vertex neighbors, leaving less headroom for $a_{l}$ to grow, which reduces the relative throughput benefits.

\paragraph{Hybrid incremental RP for GA-A is worse than its purely incremental counterparts}
RP employs a hybrid incremental approach for GA-A due to the unique requirements of the GAT architecture, i.e., once a vertex is seen in the propagation tree at hop-$l$, its embeddings for layers deeper than $l$ have to be recomputed (\S~\ref{subsubsec:attention}). As expected, we notice a significant drop in throughput for GA-A as compared to other GNNs~(Fig.~\ref{fig:results-combined-l2}, last column). E.g., with $bs=1000$, RP attains a maximum throughput of $\approx9.4\K$, $\approx2.2\K$, and $\approx1.25\K$~up/s on Arxiv, Reddit, and Products, respectively. In contrast, the maximum throughput for purely incremental workloads, GC-S -- GI-S, reaches $\approx56\K$, $\approx7.8\K$, and $\approx4\K$~up/s.

\paragraph{Hybrid incremental RP for GC-X performs relatively well despite its hybrid nature}
GC-X also uses a hybrid incremental like GA-A, but its performance on RP is competitive with purely incremental GNNs.
Despite using a hybrid approach, GC-X uses the \textit{max} function that offers a unique opportunity to prune the path of the propagation tree if a vertex's embeddings do not change. Hence, GC-X either outperforms on Products (by $1.1$--$3.4\times$) or matches for Reddit the throughput of purely incremental GNNs using RP, for $bs=1000$. But its throughput on Arxiv drops to $\approx17.6\K$~up/s, compared to $37\K$--$56\K$~up/s for the incremental models, since the fraction of active vertices at the final hop that require recomputation 
(\S~\ref{subsubsec:monotonic:3} and Fig.~\ref{fig:max-cases}(b)), is substantially higher for Arxiv~($\approx47\%$) than for Reddit~($\approx25\%$) or Products~($\approx8\%$). This increased recomputation rate significantly reduces the throughput for Arxiv.

\paragraph{The memory usage of RP is higher than the baselines, RC and DRC} 
Fig.~\ref{fig:results-combined-l2-util} reports the memory utilization of the different approaches. 
We see a noticeable increase in the memory usage of both RP and RC than DRC, increasing from $\approx43$~GiB for DRC to $\approx60$~GiB for RC and $\approx66$~GiB for RP, averaged across $bs$ and GNNs. This is due to how the graph structures are stored. While the embeddings stored in memory remain the same across the three strategies, the graph topology for DRC is stored as sparse matrices, which have a smaller memory footprint than our edge list. We also see a $\approx10\%$ uptick in RP's memory than RC due to the extra data structures to store the intermediate messages (\S~\ref{subsec:design-inc}). Lastly, we note a significant increase in the memory usage for RC and RP for GC-W~($94$~GiB for GC-W vs. $\approx62$~GiB on average for the others). This is expected since
GC-W requires RP and PC to explicitly store the edge weights in a separate data structure, as compared to DRC, where the edge weights are merged into the sparse matrix representations for the graph topology.
While RP incurs higher memory usage, this offers a significant reduction in compute time and substantial speedup in throughput, e.g., up to $128\times$ over RC for Arxiv. If memory is a limiting factor, RP can operate in a distributed setup, albeit with some performance trade-offs.

\subsubsection{Scaling with Number of GNN Layers}\label{subsec-eval-single-scalelayers}

Next, we examined the impact on performance as the number of layers in the GNN models varies.
Fig.~\ref{fig:results-products-l2-l3-l4} shows the throughput of \rpp~(RP), DRC and RC with $2$--$4$ layers. GNN models rarely exceed 4 layers. The increase in depth exponentially increases the size of each propagation tree, and can eventually encompass the whole graph.

RP sees a drop from $\approx3.1\K$ up/s to $41$~up/s as the layers increase from $2$ to $4$ for $bs=1000$. This reflects the number of affected vertices growing rapidly with the model depth, and the time taken to propagate these changes. The respective drop in RC is from $162$~up/s to $20$~up/s while DRC maintains the throughput at $~\approx1.3$~up/s. 
This is because the time to process a batch by DRC is dominated by the \textit{topology update}~(e.g., taking $\approx80\%$ for $bs=10$ for Products in Fig.~19b), which is unaffected as the number of layers increases (Appendix~\ref{appendix:A}). Hence, the throughput shows negligible change in DRC with more GNN layers; at the same time, the throughput is too low to begin with, due to the topology update costs.

For GC-W, the throughput for RC drop below that of DRC as the number of layers increases: $0.33$~up/s vs. $0.4$~up/s for $bs=1$, and $0.66$~up/s vs. $0.83$~up/s for $bs=10$. The edge list used by RC for the topology takes more time to collect the edge weights, which are maintained separately, for smaller batch sizes.
As the update time for DRC increases for larger $bs\geq100$) due to its CSR format, RC resumes outperforming DRC.

Lastly, the throughput of RP approaches RC for larger $bs=100$ and $bs=1000$ with $4$ layers. As discussed, increasing the model depth causes $a_{l-1}$ to approach $a_l$ since $a_{l-1}$ (for larger $l$) will cover a larger fraction of the graph with significant overlap in the out-neighbors
(\S~\ref{subsubsec:benefits-analysis}), leading to diminished benefits from incremental processing.

\subsubsection{Comparison with InkStream}\label{subsec-eval-single-inkvripple}
InkStream\footnote{\url{https://github.com/WuDan0399/InkStream}}~\cite{wu2025inkstream} is a recent SOTA work that is closest to us, as it too adopts an incremental inference approach to reduce redundant recomputation. We first discuss the design limitations of InkStream and then report an experimental comparison of \rpp against it.

\paragraph{Limitations of InkStream} Despite offering incremental inferencing like us, InkStream suffers from several fundamental design limitations that restrict its scalability and applicability in real-world settings. It assumes that all updates are known \textit{a priori}, i.e., even the ones that will arrive in the future, and processes them as a single batch.
To allow comparison in a streaming setting, we had to extend InkStream with new data structures that preserve the graph state across batches and invoked it iteratively.
Further, InkStream does not support attention-based models like GAT, and is
limited to only edge additions and deletions. 
Finally, InkStream lacks a distributed execution model and is limited to the memory on a single machine, e.g., preventing it from running our \textit{Papers} experiment.
\paragraph{Performance Comparison} To enable a fair comparison with InkStream~(IS), we generate a trace with equal number of just edge additions and deletions starting with the $80\%$ graph, resulting in $400\K$ events for Arxiv and $40\M$ events for Reddit and Products. As before, experiments ran for the earlier among $4$~hours or till all updates are processed.

Fig.~\ref{fig:ink-v-ripple} compares the relative performance of RP and IS on two 2-layer workloads, one purely incremental~(GC-S) and one hybrid incremental~(GC-X), evaluated for $4$ batch sizes. Across all datasets and workloads, RP achieves markedly higher throughput than IS. On Arxiv for $bs = 1000$ with GC-S, it reaches up to $74.5\K$~up/s compared to just $12\K$~up/s for IS, and with GC-X, $17.8\K$~up/s on RP against $7.7\K$~up/s for IS, with average speedups of $4.4\times$ and $2.2\times$ across all batch sizes.
On Reddit, RP offers $22.5\times$ throughput improvement over IS for GC-S and $24.0\times$ for GC-X,
while for Products, RP is $9.9\times$ and $7.0\times$ faster than IS for these GNNs

This performance gap is due to IS's inefficient data handling that is based on Python's native \textit{list} and \textit{defaultdict} within its \textit{EventQueue} for message handling, and employs dictionaries of lists to represent the graph topology. This design imposes an iterative, per-element processing model. In contrast, RP employs NumPy arrays for bookkeeping (e.g., for \textit{messages}, \textit{embeddings}), which enables efficient vectorized computation. Further, RP also supports additional aggregation operators and models like GAT, and also offers distributed execution.

\subsection{Distributed Performance}\label{subsec:eval-dist}
The distributed experiments are performed on the Papers dataset and run for a minimum of $8$~hours, or till $5\M$ updates are processed. Each partition runs on an exclusive machine, with the server running on a separate node.
For brevity, we report results for $3$ of the $7$ GNN workloads. We compare the performance of distributed \rpp~(RP) against a distributed model of recompute that we implement~(RC). As mentioned, distributed DGL~(DistDGL)~\cite{zheng2020distdgl} does not support online graph updates, while InkStream does not support distributed execution.

\begin{figure}[t]
  \centering
  \begin{minipage}[t]{0.65\linewidth}%
    \centering%
    \includegraphics[width=0.9\textwidth]{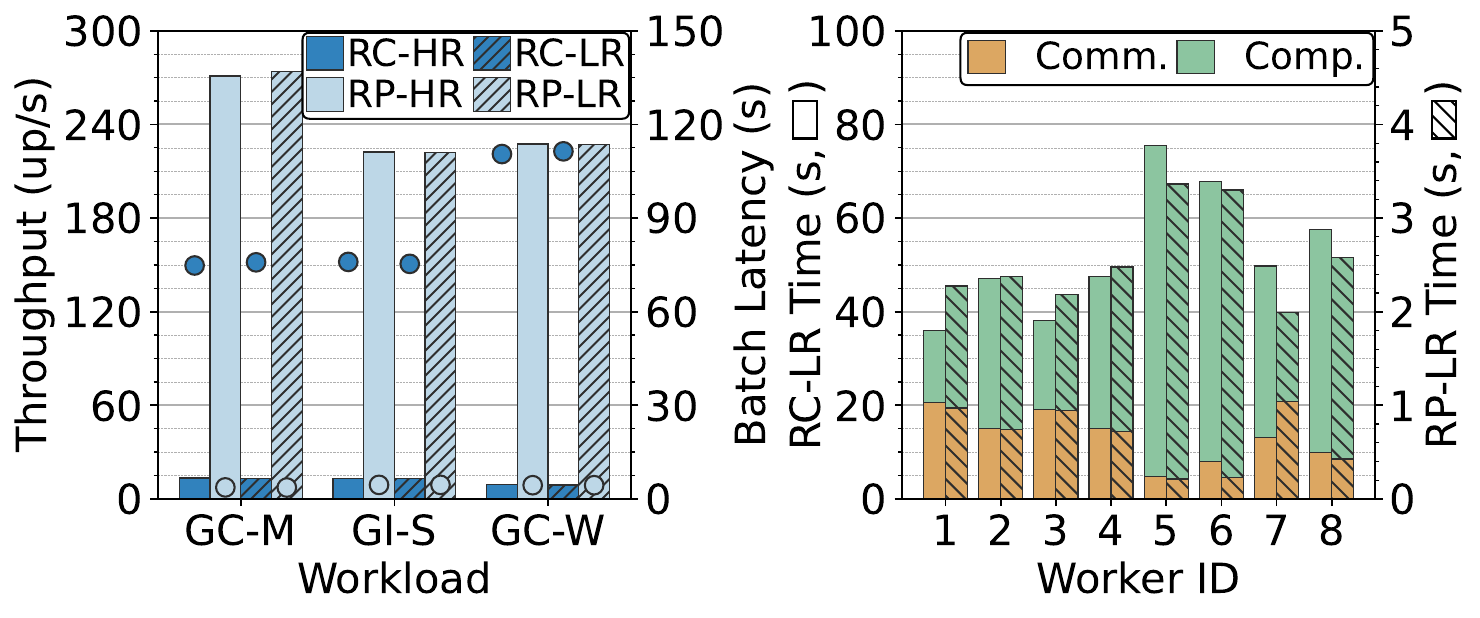}%
    \caption{Distributed performance of RP and RC on 3-layer GNNs for Papers, $bs=1000$ and $8$ partitions.}%
    \label{fig:dist-papers-8-throughput}%
  \end{minipage}~~~~
  \begin{minipage}[t]{0.35\linewidth}%
    \centering%
    \includegraphics[width=0.9\textwidth]{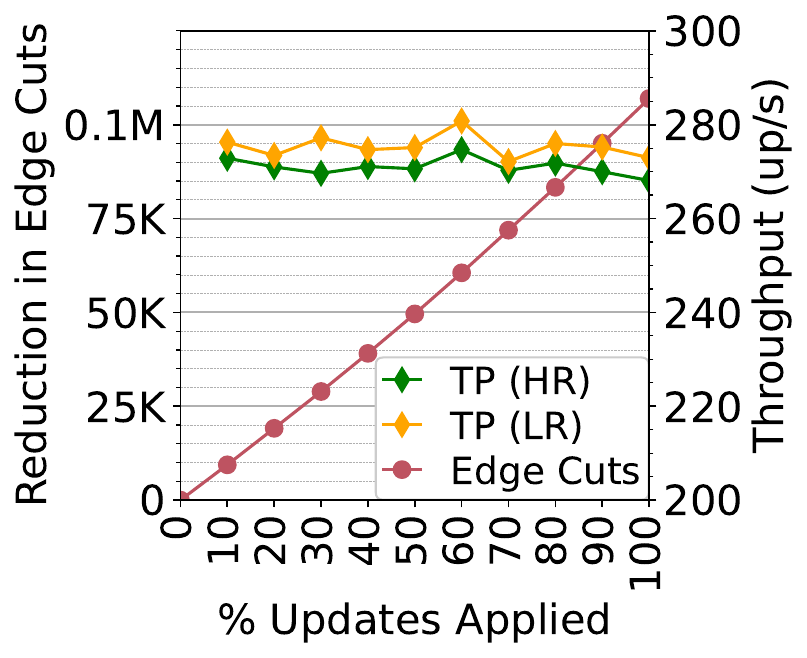}%
    \caption{Drop in edge cuts of LR vs. HR; $8$ parts, Papers, GC-M.}
    \label{fig:edge-cuts-hash-v-stream}%
  \end{minipage}%
  \vspace{-0.1in}
\end{figure}

\subsubsection{Comparison With Baseline}\label{subsec:eval-dist-baselines}
Fig.~\ref{fig:dist-papers-8-throughput} shows the throughput for GC-M, GI-S, and GC-W on Papers for $8$ partitions and $bs=1000$. RP with the hash routing achieves $20.3\times$, $16.9\times$ and $25.2\times$ faster throughput than RC for GC-M, GI-S, and GC-W, respectively~(Fig.~\ref{fig:dist-papers-8-throughput}, left subfigure, solid bar); this improvement is $20.8\times$, $16.7\times$, and $25.3\times$ using the locality-aware~(hatched bar). This confirms that RP offers much better distributed performance than RC, and achieves a peak of $270$--$274$~up/s 
for GC-M, depending on the routing.

We see a drop in the absolute throughput values for GI-S and GC-W compared to GC-M at $222$~up/s and $227$~up/s for RP. 
We also notice a sharp increase in the batch latency of RC for GC-W, from $\approx75$~seconds for GC-M and GI-S to $\approx110$~seconds for GC-W.
RP undergoes a similar, although modest, rise from $3.6$~seconds for GC-M to $4.4$~seconds for both GI-S and GC-W. This is because of the nature of the workloads. While GI-S needs to aggregate the embeddings of the vertices to the embeddings of the in-neighbors, GC-W needs to perform weighted aggregation of all the in-neighbors. The observed throughput speedup of RP over RC is a result of significant reductions in both computation and communication times per batch, e.g., a $20.5\times$ and $20\times$ drop in computation and communication times for GC-M using the streaming partitioner~(Fig.~\ref{fig:dist-papers-8-throughput}, right subfigure).

We observe that the throughput for the distributed execution is much smaller than a single-machine setup, e.g., dropping from $10.3\K$~up/s for a single-machine setup to $7\K$~up/s on $4$ workers for a 3-layer GC-M on the Arxiv graph. This is due to network, rather than in-memory, message passing, and also the need for barrier-synchronized execution required to send halo messages and perform compute at each hop.

\subsubsection{Impact of Number of Partitions}\label{subsec:eval-dist-scale}
\begin{figure}[t]
\vspace{-0.1in}
\centering
    \includegraphics[width=0.75\columnwidth]{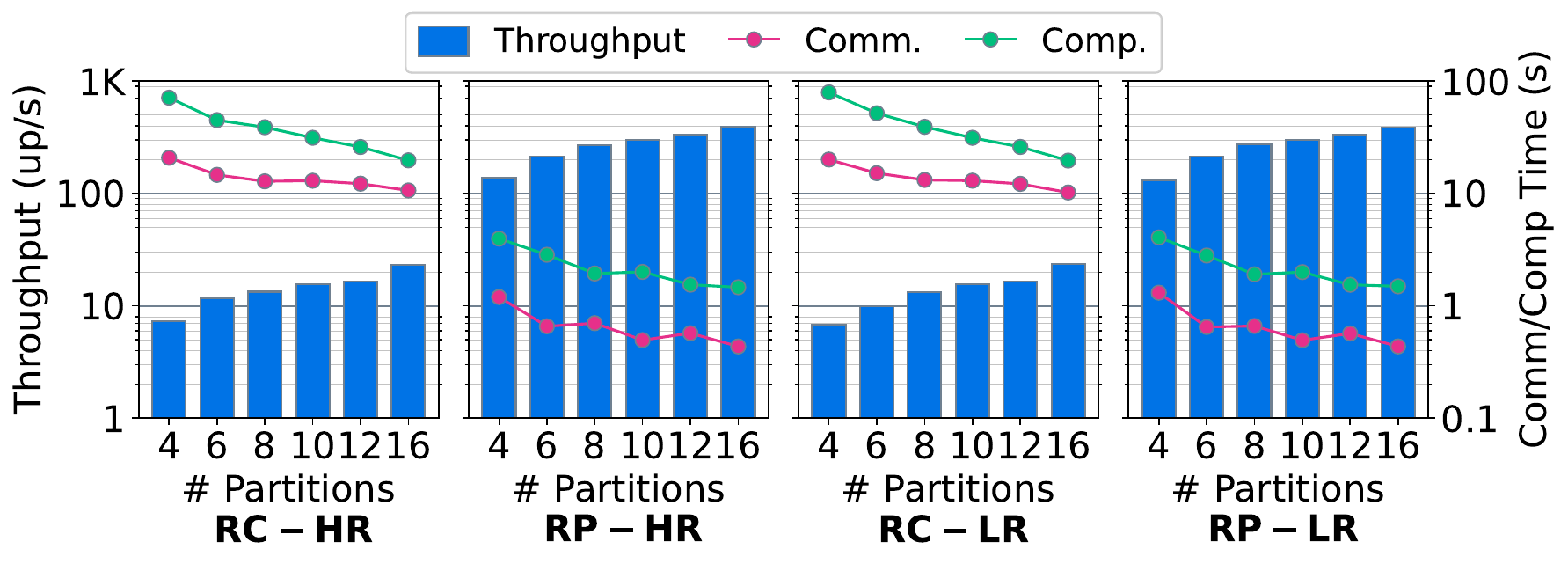}
    \caption{Distributed performance of RP and RC on 3-layer GC-M for Papers, $bs=1000$ and $4$--$16$ partitions.}
    \label{fig:dist-papers-8-scale}
    \vspace{-0.1in}
\end{figure}

Fig.~\ref{fig:dist-papers-8-scale} evaluates strong scaling of the systems with an increase in partitions from $4$ to $16$ for RP and RC, on GC-M with $bs=1000$.
As the number of partitions increases, the throughput (bar, \textit{left y-axis}, log scale) shows \textit{strong scaling} for both RC and RP. 
RC scales from $7.3$~up/s to $23.2$~up/s from $4$--$16$ partitions, with a $3.2\times$ scaling and an $80\%$ scaling efficiency. 
RP goes from $130$~up/s to $400$~up/s for $4$--$16$ partitions, which is a $3.1\times$ speedup at $77\%$ efficiency, which holds for both routing strategies. 
The scaling efficiency is ideal from $4$ to $6$ workers, at $102\%$ for RC and $109\%$ for RP, and competitive at $8$ workers ($92\%$ and $104\%$), after which it gradually reduces.
That said, the magnitude of throughput for RP is $17\times$ higher than RC.

Achieving strong scaling is not always possible in distributed systems since communication costs often increase super-linearly with the number of machines, mitigating benefits. However, in our case, as partitions increase, the batch is split into smaller batches across workers, with each taking lesser compute time to process. The compute times~(\textit{green} marker, \textit{right y-axis}, log scale) drop from $75$~seconds to $\approx19$ seconds for RC, and from $4$~seconds to $1.4$~seconds for RP as partitions grow from $4$--$16$.
At the same time, the communication time also reduces since the number of vertices that must be exchanged between workers decreases with smaller batches per worker. These drop by $\approx2\times$ for RC from $20$~seconds to $\approx10$~seconds~(\textit{pink} marker, \textit{right y-axis}, log scale) and by $\approx2.9\times$ for RP from $1.25$~seconds to $0.44$~seconds. 
RP benefits more from a drop in communications time since RC has a higher communication overhead as it pulls all remote in-neighbors at each hop. As a result, communications forms $40\%$ of its total latency compared to $29\%$ for RP.

\subsubsection{Impact of Routing Strategies}\label{subsec:eval-dist-stream}
Fig.~\ref{fig:edge-cuts-hash-v-stream} compares the performance of the hash-based (HR) and locality-aware routing (LR) strategies at the server, for the GC-M workload on Papers graph with $bs=1000$ and 8 partitions. The high-degree set is updated after every $500\K$ updates. Both strategies start with the same METIS-partitioned $80\%$ of the original graph, and the same number of edge cuts. We report the relative reduction in edge cuts for LR compared to HR (\textit{red} line, \textit{left y-axis}) after every additional $10\%$ of the $5\M$ update events are received. LR monotonically shows a steady improvement in edge cuts over HR, with about $50\K$ fewer edge cuts after 50\% of updates, and $106\K$ fewer cuts after all updates are received. In the real world, where graph sizes will continue to increase as the stream arrives, this will lead to a steadily increasing partitioning benefit for LR.

The reduction in edge cuts also leads to a better throughput for LR over HR, though it is marginal at $1.5\%$ on average (\textit{yellow} and \textit{green} lines, \textit{right y-axis}), e.g., after 60\% updates, LR has a throughput of $281$~up/s while HR has $275$~up/s.
While the lower edge cuts translate to fewer communications for LR compared to HR, it does not translate to higher throughput. 
This is because the batch latencies for both RC and RP are dominated by a \textit{straggler worker} $5$~(Fig.~\ref{fig:dist-papers-8-throughput}, right subfigure), whose execution time is compute-bound rather than communication-bound. Since we use a synchronized BSP execution, this client slows down the end-to-end execution, and mitigates edge-cut benefits for this graph.


\section{Related Work}\label{sec:related}

GNNs have found numerous applications due to their ability to learn low-level representations of graph data. E-commerce platforms like Alibaba leverage GNNs to analyze user behavior and deliver personalized product recommendations~\cite{zhu2019aligraph}. Social media platforms like Pinterest utilize GNNs for content recommendation to users~\cite{ying2018pinterest}. 
Google applies GNNs in estimating travel times for Google Maps~\cite{derrow2021google}. 
In this work, we focus on GNN applications over large dynamic graphs with millions--billions of vertices and edges, receiving streaming updates on the order of $100$--$1000$s of updates per second, as seen commonly in social, e-commerce, fintech, and road networks.

Research on GNN inference and serving has largely been limited to \textit{static graphs}~\cite{yin2022dgi,zhou2021accelerating}, where the graph structure and vertex/edge features do not change over time. Popular frameworks like DGL~\cite{wang2019deep} and Pytorch Geometric~\cite{fey2019fast} are primarily designed for GNN training and are not optimized for GNN serving. 
DGI~\cite{yin2022dgi} proposes a \textit{layer-wise} inference approach, which computes embeddings layer-by-layer and handles the tasks of all target vertices in the same layer batch-by-batch. This avoids the neighborhood explosion problem encountered while adopting \textit{vertex-wise} inference.
Frameworks such as GNNIE~\cite{mondal2022gnnie} and GraphAGILE~\cite{zhang2023graphagile} introduce specialized hardware accelerators designed to enable low-latency GNN inference, while ViTeGNN~\cite{zhou2024vitegnn} introduces an algorithm-model-architecture co-design to support low-latency GNN inference on graphs with temporal properties on FPGAs.
Hu et al.~\cite{hu2024lgrapher} propose $\lambda$-Grapher, which is a serverless system for GNN serving that achieves resource efficiency through computation graph sharing and fine-grained resource allocation for the separate memory and compute-bound requirements of a GNN layer. These methods, however, are non-trivial to extend to evolving graphs.

Real-world graphs are often dynamic in nature and require specialized frameworks. \textit{Streaming and temporal graph analytics} has been extensively studied in literature for traditional algorithms like PageRank and ShortestPath~\cite{mariappan2021dzig,vora2017kickstarter,huan2024tegraph+, besta2021practice}, and even incremental graph processing frameworks exist~\cite{taris}. Applications like PageRank on streaming graphs negate the influence of a vertex's previous state before propagating its updated state in each iteration. However, traditional solutions for streaming graph analytics are not directly applicable to GNN inference~\cite{wu2025inkstream} since GNNs handle large hidden states, making computation and memory footprint more demanding. Also, graph updates can extend to $L$-hops per update rather than being limited to immediate neighbors. This necessitates specialized approaches to streaming graph GNN inference.

There is preliminary work on GNN processing for \textit{streaming graphs}.
DGNN~\cite{yao2020dygnn} proposes an LSTM-inspired architecture to capture the dynamic nature of evolving graphs, but is limited to GNN training. OMEGA~\cite{kim2025omega} proposes selective recomputation of embeddings using a heuristic to minimize the approximation errors due to stale precomputed embeddings. They also offer parallelism strategies for graph calculations to balance inference latency and accuracy. 
Helios~\cite{jie2025helios} solves \textit{request-based} low-latency online inference for GNNs by pre-sampling during graph updates and maintaining a query-aware sample cache to reduce the sampling overhead at inference time.
STAG~\cite{wang2023stag} also solves the \textit{request-based} inference problem using a collaborative serving mechanism that balances the inference and staleness latencies.
In contrast, \rpp does \textit{trigger-based} inference while performing \textit{exact}~(not approximate), \textit{full neighborhood} embeddings update.

A recent work, InkStream~\cite{wu2025inkstream}, is an incremental framework for GNN inference on dynamic graphs that also attempts to avoid redundant computation by ``propagating only when necessary'' and ``fetching only the necessary''. Its event-based design enables pruning of update propagation, and for models with monotonic aggregation functions~(e.g., \textit{max}, \textit{min}), it reduces the update scope by focusing on the affected region instead of the entire $l$-hop neighborhood. However, it suffers from several limitations. It lacks support for attention-based architectures and restricts updates to only edge additions and deletions. Also, it assumes that all updates are available \textit{a priori} and processes them as a single batch, which is unrealistic in a streaming setting. It does not have a distributed execution model, limiting scalability when graphs do not fit on one machine. Our \rpp approach overcomes all these limitations, and, as we demonstrate, out-performs InkStream.

Further, none of these methods evaluate their solutions on large graphs with millions of vertices and $100\M+$ edges. So their scalability on single machines or in a distributed setting remains unknown. The largest public graph for GNNs, Papers~\cite{hu2020open}, is itself modest in size compared to web-scale graphs that can exceed the RAM of a single machine when considering their features and embeddings. This highlights the scalability challenges for GNN inference over large graphs.


\section{Conclusions}\label{sec:conclusion}
In this work, we present a novel framework, \rpp, to efficiently perform GNN inference over large-scale streaming graphs. Unlike traditional methods that rely on exhaustive look-back computations, our approach adopts a strictly look-forward incremental computation, where vertices are first-class entities that manage their data and updates. \rpp is able to process $\approx56\K$~up/s for sparser graphs like Arxiv and up to $7.6\K$~up/s for larger and denser graphs like Products on a single machine. Its distributed variant supports larger graphs and exhibits strong scaling, albeit at a lower throughput. As future work, we plan to explore other streaming partitioning and routing strategies to maintain balanced partitions as the graph changes, and support dynamic batch sizes for latency-sensitive tasks. Lastly, we also intend to explore how \rpp's incremental strategy could be extended to an approximate inference setup where all neighbor aggregation is avoided, in return for improved performance.


\section*{Acknowledgments}\label{sec:acknowledgements} The authors thank Roopkatha Banerjee, Tejus C., Prashanthi S.K., and other members of the DREAM:Lab, Indian Institute of Science, for their assistance and insightful feedback.

\clearpage
\bibliographystyle{plain}
\bibliography{arxiv.r0.refs}
\clearpage
\clearpage

\section{Appendix}\label{appendix}
Here, we provide additional results and discussions supporting the main text. In particular, we present detailed experiments demonstrating that our \rc and \drc implementations serve as strong and fair baselines for \rpp, along with comparisons with other DGL-based baselines on both CPU and GPU.

\subsection{Comparison with Vertex- and Layer-wise Inference}\label{appendix:A}
\begin{figure}[h]
    \centering
    \subfloat[Arxiv\label{subfig:gpu-analysis-arxiv}]{\includegraphics[width=.2\linewidth]{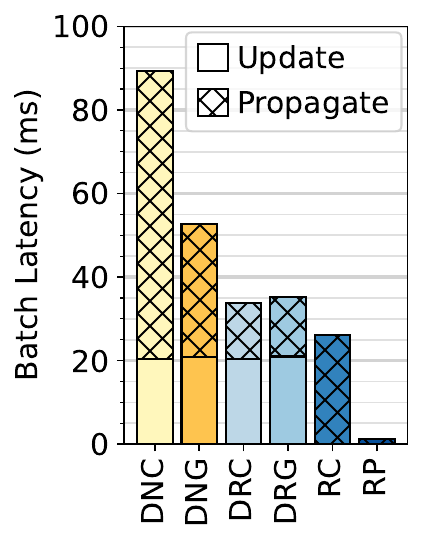}}\qquad\qquad
    \subfloat[Products\label{subfig:gpu-analysis-products}]{\includegraphics[width=.2\linewidth]{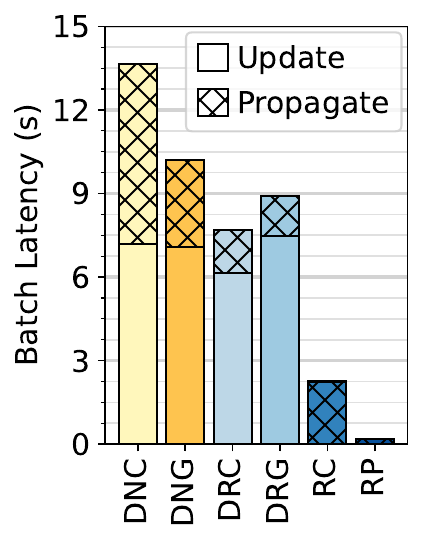}}
    \caption{Comparison of DGL's vertex- and layer-wise recompute, our layer-wise recompute~(RC), and \rpp's~(RP) incremental compute on CPU and CPU+GPU for $3$-layer GC-S and $bs=10$ on Arxiv and Products graphs.}
    \vspace{-0.1in}
    \label{fig:gpu-analysis}
\end{figure}

We first compare DGL's vertex-wise inference on the CPU~(\dnc) and CPU+GPU~(\dng), with its layer-wise recompute strategy on the CPU~(\drc) and CPU+GPU~(\drg). For the CPU+GPU configurations, the computation graphs are built on the CPU while the forward pass of inference takes place on the GPU, as is standard practice~\cite{yuan2024comprehensive,zheng2020distdgl,wang2019deep}. Further, we also contrast these against our implementation of layer-wise recompute~(\rc) and the \rpp incremental strategies~(\textbf{RP}) run on the CPU. All of these execute on the GPU workstation using a single-machine execution model.

Fig.~\ref{fig:gpu-analysis} shows the median batch latency of processing $100$ batches of $10$ updates each by these strategies for the Arxiv and Products graphs. The vertex-wise inference strategies DNC and DNG, compared against their corresponding layer-wise recompute strategies DRC and DRG are significantly slower, by $\approx2.6\times$ and $\approx1.8\times$ on the CPU, $\approx1.5\times$ and $\approx1.12\times$ on the CPU+GPU, for Arxiv and Products, respectively. This is because of the redundant computations that are performed for each batch in vertex-wise inference. Also, the GPU-based DRG method offers limited benefits over the CPU-based DRC strategy. Since the layer-wise strategy processes embeddings one layer at a time, it only requires data for the immediate previous layer in the computation graph. 
Due to these small batch sizes, the computational workload is minimal, which limits the performance gains from GPU-based computation, leading to similar performance for Arxiv, and $\approx1.1\times$ slower for Products. This justifies our choice of a layer-wise approach and just using CPUs.

Further, our own custom layer-wise recompute implementation on CPU~(RC) is $\approx1.4$--$4\times$ faster than both the CPU and GPU versions of DGL's layer-wise recompute~(DRC, DRG). 
While DGL's graph APIs ease the development of GNN training models, they are not optimized for handling a stream of updates. DGL uses sparse matrix formats for its internal graph representation. It accepts graph structures as COO matrices but internally compiles them into more efficient CSR/CSC formats for computation. Since DGL's internal structures are largely immutable, graph updates cause the entire graph structure to be rebuilt for every update~\footnote{\url{https://github.com/dmlc/dgl/blob/master/python/dgl/heterograph.py}}. Moreover, multiple types of updates in a batch can lead to multiple graph updates within a batch.
As a result, updating the graph topology consumes a significant amount of time~(\textit{Update} stack in Fig.~\ref{fig:gpu-analysis}). In contrast, our RC implementation uses lightweight edge-list structures designed to efficiently handle streaming updates. They offer much faster update times and comparable or slightly slower than the compute propagation times. The incremental computation of \rpp~(RP) is substantially faster than all of these.
In the rest of the experiments, we use DRC and RC running on the CPU as the competitive baselines to compare against \rpp's incremental approach.

\balance
\end{document}